\documentclass{article}

\usepackage[preprint]{neurips_2024}

\usepackage[utf8]{inputenc}
\usepackage[T1]{fontenc}
\usepackage{hyperref}
\usepackage{url}
\usepackage{booktabs}
\usepackage{amsmath}
\usepackage{graphicx}
\usepackage{microtype}
\usepackage{tikz}
\usetikzlibrary{shapes.geometric, arrows.meta, positioning, fit, backgrounds, calc}

\title{TechRAG: Evidence-Gated Multimodal Agentic RAG for Technical Literature Reasoning}

\author{
  Kanwar Bharat Singh \\
  Global Tire Intelligence and Solutions (GTIS) \\
  The Goodyear Tire \& Rubber Company \\
}

\begin{document}

\maketitle

\begin{abstract}
This paper presents an agentic multimodal retrieval-augmented generation (RAG) framework for domain-specific literature reasoning, instantiated on a curated corpus of several thousand papers in intelligent tires, vehicle dynamics, vehicle control, sensing, estimation, and machine learning. Unlike conventional single-pass RAG systems, the proposed architecture uses an autonomous, evidence-gated pipeline that classifies query intent, generates separate text and visual query rewrites, performs hybrid text retrieval with FAISS and BM25 followed by cross-encoder reranking, expands evidence through graph-guided chunk traversal over a Neo4j knowledge graph, and retrieves visual document evidence using ColSmol late-interaction embeddings with MUVERA fixed-dimensional encoding, approximate nearest-neighbor search, and MaxSim reranking. The framework scores evidence sufficiency using a 100-point rubric with hybrid rule-based/LLM review, retries retrieval through drift-guarded reformulation, searches external academic databases through optimize--search--vet loops, merges and deduplicates multimodal evidence, verifies citation integrity, and generates cited answers through Planner, Researcher, Writer, and Critic agents with self-correcting revision. Key contributions include: (i)~a scalable multimodal retrieval architecture combining text, graph, and visual evidence over 40,000 document pages; (ii)~an interpretable evidence sufficiency and retry mechanism; (iii)~a multi-agent generation pipeline with evidence mapping and critic-driven revision; (iv)~a domain knowledge graph with LLM-based entity extraction, OpenAlex author validation, and intra-corpus citation resolution; and (v)~a route-dependent external search architecture for targeted literature expansion. The result is a practical, evidence-gated, multimodal agentic RAG architecture for technical reasoning over specialized research corpora.
\end{abstract}

% ============================================================================
\section{Introduction}
\label{sec:introduction}
% ============================================================================

Recent advances in large language models (LLMs), enabled by the Transformer architecture~\cite{vaswani2017attention} and scaled through models such as GPT-3~\cite{brown2020language}, have demonstrated significant potential to accelerate knowledge discovery, synthesis, and reasoning across a wide range of scientific and engineering domains. Techniques such as chain-of-thought prompting~\cite{wei2022chain, kojima2022large} and tree-of-thought reasoning~\cite{yao2023tree} have further enhanced the reasoning capabilities of these models. By enabling natural-language interaction with complex information, LLMs offer new opportunities to streamline research workflows and support technical decision-making. However, when applied to highly specialized technical fields, general-purpose LLMs exhibit fundamental limitations related to domain specificity, knowledge freshness, context scalability, and traceability.

Over the past fifteen years, the author has been actively engaged in research and applied development across domains including intelligent tire systems, vehicle state estimation, vehicle dynamics, and vehicle control. This sustained involvement has necessitated continuous interaction with a broad and evolving body of specialized technical literature. As a natural outcome of this long-term research activity, the author has accumulated and curated a personal technical corpus comprising several thousand academic papers, yielding tens of thousands of indexed text chunks and approximately 40,000 document pages with visual content. The collection spans topics such as tire--road interaction modeling, intelligent tire sensing, vehicle and tire state estimation, sensor fusion, and advanced vehicle control algorithms, and has served as a foundational reference base supporting ongoing research and development efforts.

While this curated collection represents a rich and authoritative knowledge base, its growing scale and interdisciplinary nature introduce significant practical challenges for effective utilization. Traditional keyword-based search tools and manual literature review processes become progressively less effective as corpus size and technical diversity increase. Relevant prior art, methodological insights, and cross-domain relationships are often obscured by inconsistent terminology, fragmented publication venues, and the cognitive burden associated with navigating large volumes of dense technical material. Critically, much of the valuable information in technical papers is conveyed through figures, charts, diagrams, and tables that are invisible to text-only retrieval systems. As a result, valuable knowledge may remain underutilized, despite being present within the available literature.

The target users of such a system are engineers, researchers, and technical leaders working in research-heavy product development who need to connect technical literature and emerging research to generate evidence-backed insights, clarify the state of the art, drive design decisions, and support technical discussions. For these users, the ability to rapidly navigate a large body of literature with confidence in source attribution is essential.

The motivation of this work is therefore to develop a scalable, domain-specific, multimodal technical knowledge engine capable of:
\begin{itemize}
  \item Systematically organizing and indexing large volumes of specialized literature across both text and visual content
  \item Enabling semantic, meaning-based retrieval beyond conventional keyword search, including retrieval of relevant figures, charts, and diagrams directly from document pages
  \item Rapidly surfacing relevant prior art, methods, and insights across adjacent technical domains
  \item Autonomously evaluating evidence sufficiency and dynamically supplementing from external academic sources when internal evidence is insufficient
  \item Maintaining structured knowledge representations through a graph database capturing relationships between papers, authors, methods, topics, and citations
  \item Generating high-quality, cited answers through a multi-agent pipeline with structured planning, evidence mapping, and critic-driven self-correction
  \item Supporting research, system design, and technical decision-making in intelligent tire and vehicle dynamics applications
\end{itemize}

Standard RAG pipelines typically assume that retrieved chunks are sufficiently relevant once selected by the retriever. In technical literature reasoning, this assumption is fragile because relevant evidence may be incomplete, terminology may vary across subdomains, and retrieval confidence alone does not indicate whether the evidence can support an answer. Furthermore, standard text-only RAG pipelines miss the rich visual content---figures, charts, tables, and diagrams---that often conveys the most critical technical information. TechRAG addresses these gaps by making evidence sufficiency an explicit control signal, incorporating visual document retrieval alongside text retrieval, and employing multi-agent generation for structured, verifiable answer synthesis.

To address these needs, this paper presents an agentic multimodal retrieval-augmented generation (RAG) framework tailored for domain-specific technical reasoning support. Drawing on advances in agentic LLM architectures that interleave reasoning with action~\cite{yao2023react} and learn from prior attempts through self-reflection~\cite{shinn2023reflexion}, multi-agent collaboration frameworks~\cite{wu2024multiagent, hong2024metagpt}, and late-interaction visual document retrieval~\cite{faysse2025colpali}, the proposed system goes beyond conventional single-pass RAG to employ a multi-step agentic pipeline. The pipeline autonomously classifies queries by intent, generates separate text and visual retrieval queries, performs hybrid text retrieval with cross-encoder reranking, expands retrieval via graph-guided chunk traversal, retrieves visually relevant document pages using ColSmol embeddings with MUVERA-based scalable search~\cite{dhulipala2024muvera}, scores evidence sufficiency, performs agentic retry with drift-guarded reformulation, searches external academic databases through iterative loops, merges multimodal evidence, verifies citations, and generates answers through a Planner--Researcher--Writer--Critic multi-agent pipeline. By grounding model outputs in explicitly retrieved and vetted evidence from text, visual, graph, and external sources, the framework enables more reliable, transparent, and verifiable technical analysis.

The central contribution of TechRAG is not a new retrieval model, foundation model, or graph algorithm, but a practical, evidence-gated, multimodal agentic architecture with interpretable control logic, domain-specific graph augmentation, and multi-agent generation that coordinates hybrid retrieval, visual document retrieval, graph traversal, external academic search, citation verification, and self-correcting multi-agent answer generation into a reproducible workflow for technical literature reasoning. This paper makes the following contributions:
\begin{itemize}
  \item A multimodal retrieval architecture that combines hybrid text retrieval (FAISS + BM25 + cross-encoder reranking), graph-guided chunk expansion via Neo4j topic and method node traversal, and visual document retrieval using ColSmol late-interaction embeddings with MUVERA fixed-dimensional encoding for scalable search over ${\sim}40{,}000$ document pages.
  \item A multi-dimensional evidence sufficiency scoring framework (100-point rubric across five dimensions) with relevance damping, hybrid rule-based/LLM review, and agentic retry with drift-guarded reformulation, enabling the system to autonomously decide whether local evidence is sufficient or external academic sources are needed.
  \item A multi-agent answer generation pipeline with four specialized agents---Planner (structured outline), Researcher (evidence mapping with strength ratings), Writer (cited draft), and Critic (evaluation with revision loop)---replacing single-model generation with structured, verifiable synthesis.
  \item A Neo4j knowledge graph constructed via LLM-based entity extraction and OpenAlex author validation, capturing paper--author--topic--method--citation relationships with intra-corpus citation resolution, enabling co-citation traversal and structured relational retrieval.
  \item A route-dependent external search architecture with iterative optimize--search--vet--reformulate loops across Crossref, OpenAlex, and Semantic Scholar, with strategies adapted for content, bibliometric, trend, and current-world queries.
\end{itemize}

The remainder of this paper is organized as follows. Section~\ref{sec:related} surveys related work. Section~\ref{sec:limitations} discusses the fundamental limitations of standalone large language models in domain-specific settings. Section~\ref{sec:motivation} motivates the adoption of retrieval-augmented generation. Sections~\ref{sec:architecture} and~\ref{sec:implementation} describe the proposed agentic system architecture and implementation considerations, respectively. Section~\ref{sec:results} presents evaluation results. Section~\ref{sec:limitations_discuss} discusses limitations, followed by concluding remarks in Section~\ref{sec:conclusion}.

% ============================================================================
\section{Related Work}
\label{sec:related}
% ============================================================================

\paragraph{Retrieval-Augmented Generation.}
Lewis et al.~\cite{lewis2020retrieval} introduced RAG by combining a pre-trained retriever with a seq2seq generator, demonstrating that grounding generation in retrieved passages improves factuality on knowledge-intensive tasks. Dense Passage Retrieval (DPR)~\cite{karpukhin2020dense} showed that learned bi-encoder representations substantially outperform BM25 for open-domain question answering. Subsequent surveys~\cite{gao2024retrieval} have catalogued the rapid evolution of RAG architectures along the dimensions of retrieval, augmentation, and generation. TechRAG builds on this foundation but differs in three respects: it fuses dense and sparse retrieval via Reciprocal Rank Fusion~\cite{cormack2009rrf} with cross-encoder reranking~\cite{nogueira2019passage}, it augments the retriever with both a structured knowledge graph and visual document retrieval, and it replaces single-model generation with a multi-agent pipeline.

\paragraph{Self-Correcting and Adaptive RAG.}
Self-RAG~\cite{asai2024selfrag} trains a single LM to emit special reflection tokens that decide when to retrieve, whether retrieved passages are relevant, and whether the generated output is supported---enabling on-demand retrieval without a separate scoring module. Corrective RAG (CRAG)~\cite{yan2024corrective} introduces a lightweight retrieval evaluator that classifies documents as correct, ambiguous, or incorrect, and triggers web search when confidence is low. REPLUG~\cite{shi2024replug} treats the retriever as a plug-in to a frozen black-box LM and tunes retrieval via the LM's perplexity signal. TechRAG shares the adaptive retrieval philosophy of these systems but implements it through an explicit evidence sufficiency scoring rubric (100-point, five-dimension) with rule-based scoring, relevance damping, and LLM-based review, rather than through learned reflection tokens or perplexity-based signals. This design choice trades end-to-end learnability for interpretability and auditability, which are priorities in engineering domains where users need to understand why the system sought additional evidence.

\paragraph{Agentic and Multi-Step Reasoning.}
ReAct~\cite{yao2023react} demonstrated that interleaving reasoning traces with tool-use actions enables LLMs to solve multi-step tasks more reliably than reasoning or acting alone. Reflexion~\cite{shinn2023reflexion} extended this idea by adding verbal self-reflection, where an agent critiques its own failures and carries forward lessons to subsequent attempts. Tree of Thoughts (ToT)~\cite{yao2023tree} introduced deliberate search over intermediate reasoning steps. IRCoT~\cite{trivedi2023interleaving} interleaves retrieval with chain-of-thought reasoning for multi-hop questions. TechRAG draws on both ReAct-style action sequencing and Reflexion-style retry (the agentic retry uses retrieval shortcomings to guide query reformulation), but wraps these mechanisms in a bounded pipeline with explicit drift guards and accept/reject gates rather than an open-ended agent loop.

\paragraph{Multi-Agent LLM Systems.}
Recent work has demonstrated that decomposing complex tasks across multiple specialized LLM agents can outperform monolithic single-model approaches. AutoGen~\cite{wu2024multiagent} provides a framework for multi-agent conversation where agents with different roles collaborate to solve tasks. MetaGPT~\cite{hong2024metagpt} assigns distinct professional roles (architect, engineer, reviewer) to agents and coordinates them through structured outputs. TechRAG adopts a similar philosophy for answer generation: rather than relying on a single LLM call to produce a cited answer from evidence, it decomposes generation into four specialized agents---Planner, Researcher, Writer, and Critic---each with distinct roles, temperature settings, and output contracts. The Critic agent enables a self-correcting revision loop analogous to code review in MetaGPT, but applied to evidence-grounded technical writing.

\paragraph{Graph-Enhanced Retrieval.}
GraphRAG~\cite{edge2024graphrag} constructs a hierarchical community graph from source documents and uses LLM-generated community summaries to answer global, corpus-level queries that standard vector retrieval handles poorly. TechRAG takes a complementary approach: rather than summarizing communities, it builds a typed entity graph (papers, authors, topics, methods, metrics, applications) with LLM-based extraction, OpenAlex author validation, and intra-corpus citation resolution. At query time, the graph is used at two pipeline points: graph-guided chunk expansion discovers papers sharing topic or method nodes with retrieved chunks and merges qualifying chunks into the retrieval results; knowledge graph context enriches the augmented prompt with paper metadata, co-citation analysis, and author lookup. GraphRAG uses community summaries rather than retrieval expansion; TechRAG directly augments the retrieved evidence set.

\paragraph{Visual Document Retrieval.}
ColPali~\cite{faysse2025colpali} introduced late-interaction retrieval for document pages by extending the ColBERT~\cite{khattab2020colbert} multi-vector approach to vision-language models: each document page is encoded as a set of patch embeddings, and query--page relevance is computed via MaxSim (maximum similarity) scoring. This eliminates the need for OCR pipelines and enables retrieval of visually rich content (figures, tables, charts) that text extraction misses. However, the multi-vector representation creates storage and search scalability challenges---each page produces ${\sim}1{,}000$ patch vectors. MUVERA~\cite{dhulipala2024muvera} addresses this by converting multi-vector representations into fixed-dimensional encodings (FDEs) suitable for standard single-vector approximate nearest-neighbor (ANN) search. TechRAG combines ColSmol (a 500M-parameter variant optimized for limited VRAM) with MUVERA FDEs indexed in FAISS for a two-stage visual retrieval pipeline: FAISS ANN search over ${\sim}40{,}000$ page FDEs narrows candidates to 100, then full MaxSim late-interaction reranking scores only those 100 pages, achieving sub-second visual retrieval without loading the full 20~GB embedding corpus into memory.

\paragraph{Positioning.}
Table~\ref{tab:comparison} summarizes the key architectural differences. The comparison is not intended to imply superiority across all dimensions---Self-RAG's learned retrieval gating and CRAG's lightweight evaluator are more principled in settings where labeled training data is available, and GraphRAG's community summarization addresses global corpus-level queries that TechRAG does not target. Rather, TechRAG occupies a different point in the design space: it integrates multimodal evidence-gated control flow, multi-source retrieval (local hybrid, visual document, external academic APIs, knowledge graph), multi-agent generation, citation verification, and self-correcting revision into a reproducible, auditable pipeline for domain-specific technical literature reasoning.

\begin{table}[t]
\caption{Architectural comparison with related systems.}
\label{tab:comparison}
\centering
\small
\begin{tabular}{@{}lcccccc@{}}
\toprule
\textbf{Feature} & \textbf{Standard} & \textbf{Self-RAG} & \textbf{CRAG} & \textbf{GraphRAG} & \textbf{ColPali} & \textbf{TechRAG} \\
                  & \textbf{RAG}      &                   &               &                   &                  & \textbf{(Ours)}       \\
\midrule
\textit{Primary design objective} & \textit{Grounded} & \textit{Learned} & \textit{Retrieval} & \textit{Corpus-level} & \textit{Visual doc.} & \textit{Domain-specific} \\
                                   & \textit{generation} & \textit{retrieval gating} & \textit{correction} & \textit{summarization} & \textit{retrieval} & \textit{lit.\ reasoning} \\
\midrule
Hybrid dense+sparse retrieval     & No   & No      & No      & No   & No  & Yes \\
Cross-encoder reranking           & No   & No      & No      & No   & No  & Yes \\
Visual document retrieval         & No   & No      & No      & No   & Yes & Yes \\
Explicit evidence scoring         & No   & Partial$^*$ & Partial$^*$ & No & No & Yes \\
Agentic retry with drift guard    & No   & No      & No      & No   & No  & Yes \\
External academic search          & No   & No      & Web     & No   & No  & Yes \\
Knowledge graph traversal         & No   & No      & No      & Yes  & No  & Yes \\
Graph-guided retrieval expansion  & No   & No      & No      & No   & No  & Yes \\
Multi-agent generation            & No   & No      & No      & No   & No  & Yes \\
Citation verification             & No   & No      & No      & No   & No  & Yes \\
Critic-driven revision loop       & No   & Yes     & No      & No   & No  & Yes \\
\bottomrule
\end{tabular}
\\[2pt]
{\footnotesize The table compares architectural coverage, not benchmark superiority; each system targets a different design objective. $^*$Self-RAG uses learned reflection tokens; CRAG uses a retrieval evaluator. Both differ from TechRAG's explicit multi-dimensional rubric.}
\end{table}

% ============================================================================
\section{Limitations of General-Purpose Standalone Large Language Models in Domain-Specific Settings}
\label{sec:limitations}
% ============================================================================

Despite their strong language generation and reasoning capabilities, standalone/pre-trained LLMs are not well-suited for domain-critical engineering applications without augmentation~\cite{gao2024retrieval}. Key limitations include the following.

\subsection{Domain Knowledge Gaps}

Pre-trained LLMs are trained on broad, general-purpose datasets and often lack deep coverage of niche, proprietary, or highly specialized technical domains. As a result, their responses may be incomplete, overly generic, or misaligned with domain-specific assumptions and constraints common in tire mechanics and vehicle dynamics.

\subsection{Knowledge Staleness}

LLMs are trained on static datasets with a fixed knowledge cutoff. Incorporating newly published research, internal reports, or evolving best practices requires retraining or fine-tuning, which is computationally expensive and operationally complex. This makes standalone LLMs ill-suited for fast-moving research environments.

\subsection{Limited Context Window}

Large language models can process only a finite number of input tokens per inference, which fundamentally constrains their ability to reason over large documents or extensive collections of technical material. This limitation is intrinsic to transformer-based architectures~\cite{vaswani2017attention}, where both computational cost and attention complexity scale with context length.

Even with extended-context models (e.g., on the order of 32k tokens), the usable input typically corresponds to only 15--20 pages of dense technical text. This is insufficient for many real-world engineering and research workflows that require simultaneous reasoning across dozens of papers, multi-chapter reports, appendices, or large patent portfolios.

Moreover, increasing context length does not linearly translate to improved reasoning performance. As the input grows, attention mechanisms exhibit degradation in effective recall and focus, particularly for information appearing earlier in the prompt. Consequently, simply ``feeding more text'' into a standalone LLM is neither scalable nor reliable for domain-critical technical analysis.

\subsection{Lack of Source Attribution and Traceability}

A critical limitation of standalone large language models is their inability to reliably provide explicit source attribution for the information they generate. LLM outputs are produced by synthesizing patterns learned during training, rather than by retrieving and citing specific documents, sections, or evidence. As a result, generated responses are typically uncited, non-verifiable, and difficult to trace back to authoritative sources.

This limitation is particularly problematic in research and development settings, where technical claims must be supported by verifiable references. Without clear provenance, it becomes challenging to assess the validity of assumptions, reproduce results, compare methodologies, or build confidence in engineering decisions derived from LLM-generated content.

% ============================================================================
\section{Motivation for Retrieval-Augmented Generation}
\label{sec:motivation}
% ============================================================================

Collectively, the limitations of standalone large language models outlined in the preceding section motivate retrieval-augmented generation (RAG) approaches~\cite{gao2024retrieval}, which decouple knowledge storage from reasoning and enable LLMs to operate over large, evolving technical corpora in a controlled, transparent, and traceable manner. The present use case is particularly well suited for a RAG architecture due to the nature, scale, and technical value of the underlying knowledge corpus.

The corpus consists of several thousand highly specialized academic papers, technical reports, and research documents focused on intelligent tires, smart tire sensing, tire--road interaction, vehicle dynamics, and vehicle control systems. This domain is characterized by dense technical content, mathematical formulations, domain-specific terminology, and a strong dependence on modeling assumptions, operating conditions, and experimental regimes.

Traditional keyword-based search is insufficient in this context for several reasons:
\begin{itemize}
  \item Relevant concepts are frequently expressed using different terminology across authors, disciplines, and time periods
  \item Critical insights are often embedded within explanatory or methodological text, rather than appearing as explicit keywords
  \item Cross-disciplinary relationships---such as the influence of tire mechanics on ABS or ESC control behavior---are difficult to uncover through lexical matching alone
  \item Figures, charts, tables, and diagrams---which often convey the most important technical content---are entirely invisible to keyword-based and standard text-embedding retrieval
\end{itemize}

At the same time, pure LLM-based approaches without retrieval are inadequate for research-grade and safety-relevant technical workflows, as they:
\begin{itemize}
  \item Lack direct access to the full domain-specific corpus
  \item Risk hallucinating or oversimplifying technical details
  \item Provide limited traceability to original sources, prior art, and supporting evidence
\end{itemize}

Retrieval-augmented generation directly addresses these limitations by:
\begin{itemize}
  \item Enabling semantic, meaning-based retrieval across the entire corpus
  \item Grounding generated responses in retrieved, authoritative source documents
  \item Preserving transparency by allowing inspection and verification of supporting text
  \item Scaling efficiently to thousands of documents through dense vector indexing
\end{itemize}

However, basic single-pass RAG systems---which retrieve a fixed set of text chunks and pass them directly to a language model---remain limited in their ability to handle the diversity of query types and information modalities encountered in research workflows. A content question about tire modeling requires different retrieval strategies than a bibliometric query about publication trends. Furthermore, a single retrieval pass may produce insufficient or weakly relevant evidence with no mechanism to detect or recover from such failures. And critically, text-only retrieval misses the visual evidence---figures, charts, and tables---that often conveys the most important technical information in engineering papers.

These observations motivate the evolution from basic RAG to an agentic multimodal RAG architecture. Recent work on agentic LLM frameworks has demonstrated that interleaving reasoning with actions such as tool use and search~\cite{yao2023react} and incorporating self-reflection~\cite{shinn2023reflexion} substantially improve task performance. Late-interaction visual document retrieval~\cite{faysse2025colpali} has shown that vision-language models can retrieve document pages by visual content without OCR. Multi-agent collaboration~\cite{wu2024multiagent, hong2024metagpt} has demonstrated that decomposing complex tasks across specialized agents improves output quality. Drawing on these advances, the proposed system autonomously classifies query intent, retrieves evidence from text, visual, and graph modalities, evaluates evidence quality, performs iterative retrieval with drift-guarded reformulation, searches external academic databases when needed, and generates answers through a multi-agent pipeline with critic-driven revision. The following section presents this agentic architecture in detail.

% ============================================================================
\section{System Architecture and Methodology}
\label{sec:architecture}
% ============================================================================

The proposed agentic multimodal RAG system is a modular, research-grade technical knowledge engine whose architecture is domain-agnostic, while its instantiation in this work is tailored to large-scale, domain-specific literature in intelligent tires, vehicle state estimation, vehicle dynamics, and vehicle control. The system is explicitly designed to address challenges related to multimodal semantic retrieval, context scalability, evidence sufficiency, and reasoning faithfulness in safety-relevant engineering domains.

The pipeline comprises orchestrated stages that execute autonomously for each user query, organized into two agentic loops (Figure~\ref{fig:architecture}). \textbf{Loop~1 (Multimodal Evidence Gathering)} encompasses query classification with text and visual query rewriting, hybrid text retrieval, graph-guided chunk expansion, visual document retrieval (content route only), evidence sufficiency scoring with agentic retry, and external academic search when evidence is insufficient. Its objective is to assemble the strongest possible multimodal evidence set from text, visual, graph, and external sources. \textbf{Loop~2 (Multi-Agent Answer Curation)} takes the assembled evidence and performs evidence merging with deduplication and visual page attachment, multimodal prompt construction, citation verification, multi-agent answer generation (Planner $\to$ Researcher $\to$ Writer $\to$ Critic), and critic-driven revision. This separation ensures that evidence quality is established before synthesis begins, and that the synthesis itself is independently verified through a structured multi-agent process.

A defining characteristic of the architecture is a layered guardrail design with three quality gates at distinct pipeline stages:
\begin{enumerate}
  \item \textbf{Evidence Sufficiency Guardrail} (Step~5, post-retrieval): Scores the retrieved evidence against a 100-point rubric with LLM review. Controls whether the pipeline proceeds with local evidence, triggers agentic retry, or activates external search. A drift guard prevents query reformulation from straying off-topic.
  \item \textbf{Citation Verification Guardrail} (Step~9, pre-generation): After the full multimodal prompt is assembled---with text chunks, external abstracts, graph context, and visual metadata---an LLM checks whether the assembled sources actually cover the question, flags gaps and contradictions, and appends verification notes to the prompt so the writer agents can account for identified weaknesses.
  \item \textbf{Critic Guardrail} (Step~10, post-generation): After the writer produces a cited draft, the critic agent evaluates citation correctness, caveat inclusion, unsupported claims, and hallucinations. A failing evaluation triggers a targeted revision loop.
\end{enumerate}
These guardrails operate at three different levels of abstraction---retrieved chunks, assembled prompt, and generated answer---ensuring that quality is assessed at each stage rather than relying on a single end-to-end check.

All computationally intensive preprocessing steps---including document parsing, structure-aware chunking, text embedding generation, visual page embedding with ColSmol, MUVERA FDE computation, FAISS index construction, entity extraction, and knowledge graph construction---are performed offline.

\begin{figure}[t]
\centering
\small
\setlength{\fboxsep}{6pt}
\fbox{\parbox{0.92\textwidth}{%
\textbf{Loop 1: Multimodal Evidence Gathering} \\[4pt]
\texttt{Step 1} Query Classification + Rewrite (text + visual) $\to$
\texttt{Step 2} Text Retrieval (FAISS+BM25+RRF+Rerank) $\to$
\texttt{Step 3} Graph-Guided Chunk Expansion (Neo4j) $\to$
\texttt{Step 4} Visual Retrieval (ColSmol+MUVERA+MaxSim; content route only) $\to$
\texttt{Step 5} Evidence Sufficiency Check [Guardrail~1] (rubric + LLM + agentic retry if WEAK) $\to$
\texttt{Step 6} External Evidence Loop (if MODERATE/WEAK; Crossref+OpenAlex+SS) \\[6pt]
\textbf{Loop 2: Multi-Agent Answer Curation} \\[4pt]
\texttt{Step 7} Evidence Merge (dedup + rank + attach visual pages + graph metadata enrichment) $\to$
\texttt{Step 8} Augmented Multimodal Prompt $\to$
\texttt{Step 9} Citation Verification [Guardrail~2] $\to$
\texttt{Step 10} Multi-Agent Generation [Guardrail~3: Critic] (Planner$\to$Researcher$\to$Writer$\to$Critic) $\to$
\texttt{Step 11} Final Result
}}
\caption{TechRAG pipeline architecture organized into two agentic loops with three guardrails. Loop~1 gathers and scores multimodal evidence from text, visual, graph, and external sources (Guardrail~1: evidence sufficiency). Loop~2 enriches with graph metadata, merges evidence, verifies citations (Guardrail~2), and generates the final answer through a multi-agent pipeline with critic-driven revision (Guardrail~3).}
\label{fig:architecture}
\end{figure}

\subsection{Pipeline Overview}
\label{sec:pipeline}

The pipeline stages are summarized in Table~\ref{tab:pipeline}.

\begin{table}[t]
\caption{Agentic multimodal RAG pipeline: orchestrated stages.}
\label{tab:pipeline}
\centering
\small
\begin{tabular}{@{}clp{7.5cm}c@{}}
\toprule
\textbf{Step} & \textbf{Name} & \textbf{Description} & \textbf{LLM?} \\
\midrule
\multicolumn{4}{@{}l}{\textit{Loop 1: Multimodal Evidence Gathering}} \\
1  & Query Classification + Rewrite & Classify route (content/bibliometric/trend/current\_world); rewrite for text retrieval; rewrite for visual retrieval (content route) & Yes \\
2  & Text Retrieval         & FAISS (semantic) + BM25 (keyword) + RRF fusion + cross-encoder reranking & No \\
3  & Graph Expansion        & Neo4j topic/method traversal from retrieved chunks; IDF-weighted; cross-encode related chunks; ALIAS\_OF resolution; merge qualifying results & No \\
4  & Visual Retrieval       & Content route only: ColSmol query embedding $\to$ MUVERA FDE $\to$ FAISS ANN (${\sim}40$K pages) $\to$ MaxSim rerank top-100 $\to$ top-$K$ pages & No \\
5  & Evidence Sufficiency Check & Rule-based 100-pt rubric + LLM reviewer (can only downgrade); agentic retry if WEAK (reformulate $\to$ drift guard $\to$ re-retrieve $\to$ second hybrid check) [Guardrail~1] & Yes \\
6  & External Evidence Loop & If MODERATE/WEAK: Crossref+OpenAlex+SS (3 attempts) or OpenAlex agentic loop (5 attempts); optimize $\to$ search $\to$ LLM vet $\to$ reformulate & Yes \\
\midrule
\multicolumn{4}{@{}l}{\textit{Loop 2: Multi-Agent Answer Curation}} \\
7  & Evidence Merge         & Rank, deduplicate chunks (1,800-char hash) and sentences (400-char hash); attach top-5 visual pages as base64 JPEG; format graph context as metadata & No \\
8  & Augmented Multimodal Prompt & Assemble [SOURCE $n$] text + [EXTERNAL $n$] abstracts + graph context + [VISUAL $n$] page images into structured prompt & No \\
9  & Citation Verification  & Full prompt + explicit query $\to$ LLM checks for gaps, contradictions, off-topic sources [Guardrail~2] & Yes \\
10 & Multi-Agent Generation & Planner (outline) $\to$ Researcher (evidence map) $\to$ Writer (cited draft) $\to$ Critic [Guardrail~3] (pass/revise; max 1 revision loop) & Yes \\
11 & Final Result           & Answer + citations + confidence + quality medal + multi-agent trace + visual evidence gallery & No \\
\bottomrule
\end{tabular}
\end{table}

\subsection{Query Classification and Rewriting (Step 1)}
\label{sec:classification}

Before retrieval begins, GPT-4o-mini classifies each incoming query into one or more retrieval routes and generates optimized retrieval queries. This step performs three functions:

\paragraph{Route Classification.} The classifier determines which retrieval branches and external search strategies to activate. The four routes are:
\begin{itemize}
  \item \textbf{Content}: Technical/engineering questions. Activates text retrieval, graph expansion, visual retrieval, and Crossref+OpenAlex+Semantic Scholar external search (3 attempts).
  \item \textbf{Bibliometric}: Queries seeking papers on a topic, latest publications, or author searches. Activates text retrieval, graph expansion (with author lookup), and OpenAlex agentic loop (5 attempts). Visual retrieval is skipped.
  \item \textbf{Trend}: Queries about research evolution or publication volume over time. Uses OpenAlex trend aggregation (\texttt{group\_by=publication\_year}). Visual retrieval is skipped.
  \item \textbf{Current\_world} (auxiliary): Company strategies, market data, regulations, industry news. Uses web search, bypassing academic retrieval entirely.
\end{itemize}
A query can have multiple routes (e.g., both bibliometric and content).

\paragraph{Text Query Rewrite.} GPT-4o-mini generates a semantically equivalent but technically explicit rewritten query, expanding abbreviations, normalizing phrasing toward literature-style language, and adding technical synonyms. This runs for all routes.

\paragraph{Visual Query Rewrite.} For content-route queries, a separate LLM call generates a visual retrieval query optimized for matching document page imagery---rewritten into the form ``figure or chart showing \ldots'' to align with ColSmol's visual matching capabilities. This query is used exclusively for the visual retrieval branch (Step~4).

\subsection{Structure-Aware Chunking (Offline)}
\label{sec:chunking}

\paragraph{Section-Aware Chunking.} Each document is segmented based on detected structural markers, including explicit section headers (e.g., Abstract, Introduction, Methods, Results), numbered headings, and common scholarly section conventions. Chunk boundaries are constrained to lie entirely within individual sections; chunks are never permitted to span across section boundaries.

\paragraph{Paragraph-Aware Chunking.} Within each section, text is further segmented into complete paragraphs using layout-based cues and newline heuristics. Chunk construction adheres to a strict integrity rule: only whole paragraphs are aggregated, and sentences or paragraphs are never split.

\paragraph{Chunk Sizing and Overlap.} Chunks are constructed using empirically selected parameters: target size of approximately 3,500 characters, minimum approximately 300 characters, maximum approximately 6,000 characters, and overlap of approximately 400 characters. Each resulting chunk represents a self-contained semantic unit aligned with the document's natural structure.

\paragraph{Implementation.} Document parsing and text extraction are performed using PyMuPDF~\cite{pymupdf2024}. For each chunk, metadata including document identifier, section label, page range, and chunk unique identifier is retained to support traceability, citation, and linkage to the knowledge graph.

\subsection{Embedding and Lexical Representations (Offline)}
\label{sec:embedding}

\paragraph{Dense Semantic Embeddings.} Each document chunk is embedded using \texttt{sentence-transformers/all-MiniLM-L6-v2}~\cite{reimers2019sentence}, producing a 384-dimensional dense vector. At query time, the rewritten user query is embedded using the same model.

\paragraph{Sparse Lexical Representation (BM25).} In parallel, each chunk is represented via BM25~\cite{robertson2009bm25} to support exact keyword-based retrieval. Chunk text is tokenized using a domain-aware normalization pipeline that preserves alphanumeric tokens, acronyms, and technical symbols.

\subsection{Hybrid Text Retrieval (Step 2)}
\label{sec:hybrid}

The system employs a hybrid retrieval strategy combining dense vector similarity search via FAISS~\cite{johnson2021faiss} with lexical keyword-based retrieval via BM25. Given the embedded query, FAISS returns ranked candidates based on cosine similarity; in parallel, BM25 produces a ranked list based on lexical relevance. The two lists are fused using Reciprocal Rank Fusion (RRF)~\cite{cormack2009rrf}.

\paragraph{Cross-Encoder Reranking.} Following RRF fusion, a cross-encoder model (\texttt{cross-encoder/ms-marco-MiniLM-L-6-v2})~\cite{nogueira2019passage} jointly encodes each query--chunk pair and predicts a relevance score. Unlike bi-encoder embeddings, the cross-encoder attends jointly over both inputs, enabling fine-grained matching of technical phrasing and methodological detail.

\subsection{Graph-Guided Chunk Expansion (Step 3)}
\label{sec:graph_expansion}

After text retrieval, the Neo4j knowledge graph discovers papers sharing \texttt{Topic} or \texttt{Method} nodes with the retrieved chunks but missed by vector and keyword search due to terminology mismatch. Chunks from these related papers are prioritized by section (abstract $\to$ introduction $\to$ method $\to$ results), cross-encoded against the query using the same cross-encoder model, and only those scoring $\geq 0.0$ are merged into the retrieval results. Expansion uses IDF weighting so that rare shared entities (e.g., ``magnetorheological damper'') rank higher than common ones (e.g., ``Kalman filter''). Retrieved chunks are tagged as \texttt{VECTOR} or \texttt{GRAPH} in the user interface, providing transparency about evidence provenance.

\subsection{Visual Document Retrieval (Step 4)}
\label{sec:visual}

For content-route queries, the system retrieves visually relevant document pages---figures, charts, tables, diagrams---directly from rendered PDF pages without OCR. This addresses a fundamental limitation of text-only RAG: in engineering papers, critical technical content is frequently conveyed through visual elements that text extraction cannot capture.

\paragraph{Visual Embedding Model.} Each document page is embedded offline using ColSmol-500M, a compact variant of the ColPali~\cite{faysse2025colpali} late-interaction vision-language model. ColSmol encodes each page as a set of ${\sim}1{,}031$ patch embeddings (128-dimensional each), capturing both textual and visual content at patch-level granularity. At query time, the visual query is encoded into 32 token embeddings ($32 \times 128$-dim).

\paragraph{MUVERA Fixed-Dimensional Encoding.} Storing and searching ${\sim}1{,}031$ vectors per page across ${\sim}40{,}000$ pages (${\sim}40$M vectors total, ${\sim}20$~GB) is infeasible for interactive retrieval. MUVERA~\cite{dhulipala2024muvera} addresses this by converting each page's multi-vector representation into a single fixed-dimensional encoding (FDE) of 4,096 dimensions (${\sim}16$~KB per page). The FDE is computed using SimHash-based partitioning (6 projections $\to$ 64 partitions) and count-sketch aggregation, preserving approximate late-interaction similarity in a single-vector form amenable to standard ANN search. The MUVERA FAISS index stores ${\sim}40{,}000$ page FDEs in ${\sim}660$~MB.

\paragraph{Two-Stage Retrieval Pipeline.} Visual retrieval proceeds in two stages:
\begin{enumerate}
  \item \textbf{FAISS ANN search}: The query FDE is compared against all ${\sim}40{,}000$ page FDEs via dot product, returning the top-100 candidate pages (${\sim}30$ms).
  \item \textbf{MaxSim reranking}: For only the top-100 candidates, the full multi-vector \texttt{.pt} embeddings are loaded from disk on demand (${\sim}50$~MB transient), and exact late-interaction MaxSim scoring~\cite{khattab2020colbert} is computed (${\sim}150$ms). The top-$K$ pages (typically 5) are selected.
\end{enumerate}

This two-stage design avoids loading 20~GB of embeddings into RAM while achieving sub-second total visual retrieval latency. Selected pages are rendered as JPEG images at 144~DPI via PyMuPDF and attached to the multimodal prompt as base64 content for the answer model.

\paragraph{No OCR Dependency.} Visual pages are rendered on-the-fly and sent directly to multimodal LLMs as images. The model interprets figures, tables, and charts natively, eliminating OCR error propagation.

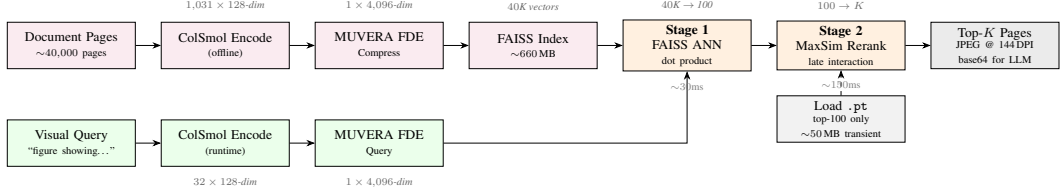
\begin{figure}[t]
\centering
\resizebox{\textwidth}{!}{%
\begin{tikzpicture}[
  stage/.style={rectangle, draw=black, fill=blue!8, text width=2.8cm, minimum height=1.1cm, align=center, font=\footnotesize},
  dim/.style={font=\scriptsize\itshape, text=black!60},
  arr/.style={-{Stealth[length=2mm]}, semithick},
  node distance=0.4cm and 0.6cm,
]
% Document path (top)
\node[stage, fill=purple!8] (pages) {Document Pages\\{\scriptsize ${\sim}40{,}000$ pages}};
\node[stage, fill=purple!8, right=of pages] (colsmol_d) {ColSmol Encode\\{\scriptsize (offline)}};
\node[stage, fill=purple!8, right=of colsmol_d] (muvera_d) {MUVERA FDE\\{\scriptsize Compress}};
\node[stage, fill=purple!8, right=of muvera_d] (faiss_idx) {FAISS Index\\{\scriptsize ${\sim}660$\,MB}};

% Dimension labels (document path)
\node[dim, above=0.15cm of colsmol_d] {$1{,}031 \times 128$-dim};
\node[dim, above=0.15cm of muvera_d] {$1 \times 4{,}096$-dim};
\node[dim, above=0.15cm of faiss_idx] {$40$K vectors};

\draw[arr] (pages) -- (colsmol_d);
\draw[arr] (colsmol_d) -- (muvera_d);
\draw[arr] (muvera_d) -- (faiss_idx);

% Query path (bottom)
\node[stage, fill=green!8, below=1.2cm of pages] (vquery) {Visual Query\\{\scriptsize ``figure showing\ldots''}};
\node[stage, fill=green!8, right=of vquery] (colsmol_q) {ColSmol Encode\\{\scriptsize (runtime)}};
\node[stage, fill=green!8, right=of colsmol_q] (muvera_q) {MUVERA FDE\\{\scriptsize Query}};

\node[dim, below=0.15cm of colsmol_q] {$32 \times 128$-dim};
\node[dim, below=0.15cm of muvera_q] {$1 \times 4{,}096$-dim};

\draw[arr] (vquery) -- (colsmol_q);
\draw[arr] (colsmol_q) -- (muvera_q);

% Stage 1: FAISS ANN
\node[stage, fill=orange!12, right=of faiss_idx] (ann) {\textbf{Stage 1}\\FAISS ANN\\{\scriptsize dot product}};
\node[dim, above=0.15cm of ann] {$40$K $\to$ 100};

\draw[arr] (faiss_idx) -- (ann);
\draw[arr] (muvera_q) -| (ann);

% Stage 2: MaxSim
\node[stage, fill=orange!12, right=of ann] (maxsim) {\textbf{Stage 2}\\MaxSim Rerank\\{\scriptsize late interaction}};
\node[dim, above=0.15cm of maxsim] {$100 \to K$};

\draw[arr] (ann) -- (maxsim);

% Load .pt annotation
\node[stage, fill=gray!10, below=0.6cm of maxsim] (pt) {Load \texttt{.pt}\\{\scriptsize top-100 only\\${\sim}50$\,MB transient}};
\draw[arr, dashed] (pt) -- (maxsim);

% Output
\node[stage, fill=black!8, right=of maxsim] (output) {Top-$K$ Pages\\{\scriptsize JPEG @ 144\,DPI\\base64 for LLM}};
\draw[arr] (maxsim) -- (output);

% Timing labels
\node[font=\scriptsize, text=black!50, below=0.15cm of ann] {${\sim}30$ms};
\node[font=\scriptsize, text=black!50, below=0.15cm of maxsim] {${\sim}150$ms};

\end{tikzpicture}
}% end resizebox
\caption{Two-stage visual retrieval pipeline. Document pages are encoded offline with ColSmol and compressed to MUVERA fixed-dimensional encodings (FDEs) for FAISS indexing. At query time, the visual query follows the same encoding path; Stage~1 narrows ${\sim}40{,}000$ candidates to 100 via ANN dot product, and Stage~2 performs exact MaxSim late-interaction reranking on only those 100 pages using on-demand \texttt{.pt} embeddings.}
\label{fig:visual_retrieval}
\end{figure}

\subsection{Evidence Sufficiency Check (Step 5)}
\label{sec:evidence}

A distinguishing feature of the proposed architecture is the autonomous assessment of whether retrieved evidence is sufficient to answer the user's question. The check comprises three stages: rule-based scoring, LLM review, and agentic retry.

\paragraph{Rule-Based Scoring.} The system scores evidence across five dimensions using a 100-point rubric:
\begin{itemize}
  \item \textbf{Retrieval Confidence} (40 points): Cross-encoder reranker scores---are the top chunks actually relevant?
  \item \textbf{Answer Specificity} (25 points): Do chunks contain specific data, methods, equations, results?
  \item \textbf{Source Diversity} (15 points): How many distinct papers contribute relevant evidence?
  \item \textbf{Metadata Completeness} (10 points): Do chunks have section labels, page numbers, extractable year?
  \item \textbf{Recency / Intent Fit} (10 points): Are sources recent enough given the query type?
\end{itemize}

\paragraph{Relevance Damping.} When retrieval confidence is low, the other four dimensions are scaled down proportionally:
\begin{equation}
  \text{damping} = \max\!\left(\min\!\left(\frac{\text{retrieval\_score}}{25},\; 1.0\right),\; 0.2\right)
\end{equation}
This prevents metadata or recency scores of off-topic chunks from inflating the overall score.

\paragraph{LLM Reviewer (Hybrid Check).} After rule-based scoring, GPT-4o-mini reviews evidence snippets and can downgrade (never upgrade) the verdict if the chunks do not actually address the question.

\paragraph{Agentic Retry.} When evidence is WEAK, the system attempts a smarter search before resorting to external sources, drawing on the principle of verbal self-reflection~\cite{shinn2023reflexion}:
\begin{itemize}
  \item \textbf{Query Reformulation}: GPT-4o-mini generates alternative search terms based on what concepts are missing from the initial retrieval.
  \item \textbf{Drift Guard}: A rule-based check ensures reformulated results still relate to the original question ($\geq 30\%$ term overlap required). If drift is detected, the retry is rejected.
  \item \textbf{Second Hybrid Check}: The full hybrid evidence check (rule-based + LLM reviewer) runs on the new results.
  \item \textbf{Accept/Reject}: Retry results are used only if the new score exceeds the original.
\end{itemize}

\paragraph{Decision Thresholds.} The final verdict determines downstream behavior:
\begin{itemize}
  \item 80--100 (\textbf{STRONG}): Answer from internal corpus only. External search is skipped.
  \item 50--79 (\textbf{MODERATE}): Answer from internal corpus, enrich with targeted external search.
  \item 0--49 (\textbf{WEAK}): Full external evidence loop (Step~6).
\end{itemize}

\begin{figure}[t]
\centering
\resizebox{\textwidth}{!}{%
\begin{tikzpicture}[
  dim/.style={rectangle, draw=black, fill=blue!8, text width=2.4cm, minimum height=0.7cm, align=center, font=\footnotesize},
  proc/.style={rectangle, draw=black, fill=orange!10, text width=2.6cm, minimum height=0.7cm, align=center, font=\footnotesize},
  dec/.style={diamond, draw=black, fill=orange!12, text width=1.2cm, align=center, font=\footnotesize, aspect=2.2},
  verdict/.style={rectangle, rounded corners=3pt, draw=black, font=\footnotesize\bfseries, minimum height=0.6cm, align=center, text width=2.2cm},
  arr/.style={-{Stealth[length=2mm]}, semithick},
  darr/.style={-{Stealth[length=2mm]}, semithick, dashed},
  node distance=0.5cm and 0.5cm,
]

% Five scoring dimensions
\node[dim] (d1) {Retrieval\\Confidence\\{\scriptsize (40\,pts)}};
\node[dim, right=0.3cm of d1] (d2) {Answer\\Specificity\\{\scriptsize (25\,pts)}};
\node[dim, right=0.3cm of d2] (d3) {Source\\Diversity\\{\scriptsize (15\,pts)}};
\node[dim, right=0.3cm of d3] (d4) {Metadata\\Completeness\\{\scriptsize (10\,pts)}};
\node[dim, right=0.3cm of d4] (d5) {Recency /\\Intent Fit\\{\scriptsize (10\,pts)}};

% Damping annotation
\draw[darr, red!60] (d1.south) -- ++(0,-0.4) -| node[below, pos=0.25, font=\scriptsize\itshape, text=red!70, text width=3cm, align=center] {relevance damping\\scales dims 2--5} (d3.south);

% Summation
\node[proc, below=1.2cm of d3] (sum) {Rule-Based Total\\{\scriptsize 0--100 points}};
\draw[arr] (d1.south) |- (sum.west);
\draw[arr] (d2.south) -- (sum);
\draw[arr] (d3.south) -- (sum);
\draw[arr] (d4.south) |- (sum.east);
\draw[arr] (d5.south) |- ([xshift=0.3cm]sum.east);

% LLM reviewer
\node[proc, fill=green!8, below=0.5cm of sum] (llm) {LLM Reviewer\\{\scriptsize can only \textbf{downgrade}}};
\draw[arr] (sum) -- (llm);

% Decision
\node[dec, below=0.7cm of llm] (dec1) {Verdict?};
\draw[arr] (llm) -- (dec1);

% Three branches
\node[verdict, fill=green!15, right=2.5cm of dec1] (strong) {STRONG\\{\scriptsize (80--100)}};
\node[verdict, fill=yellow!20, below right=0.8cm and 0.8cm of dec1] (moderate) {MODERATE\\{\scriptsize (50--79)}};
\node[verdict, fill=red!12, below=0.8cm of dec1] (weak) {WEAK\\{\scriptsize (0--49)}};

\draw[arr] (dec1) -- node[above, font=\scriptsize] {$\geq 80$} (strong);
\draw[arr] (dec1) -- node[right, font=\scriptsize, pos=0.4] {50--79} (moderate);
\draw[arr] (dec1) -- node[left, font=\scriptsize] {$< 50$} (weak);

% Actions
\node[font=\scriptsize, right=0.2cm of strong, text width=2.4cm, align=left] {Skip external\\search entirely};
\node[font=\scriptsize, right=0.2cm of moderate, text width=2.4cm, align=left] {Targeted external\\search (Step~6)};

% Agentic retry from WEAK
\node[proc, fill=red!8, below=0.5cm of weak] (retry) {Agentic Retry\\{\scriptsize reformulate $\to$\\drift guard $\to$\\re-retrieve}};
\draw[arr] (weak) -- (retry);

\node[proc, fill=red!8, right=1.2cm of retry] (recheck) {Second Hybrid\\Check\\{\scriptsize rule + LLM}};
\draw[arr] (retry) -- (recheck);

\node[dec, right=1.0cm of recheck] (accept) {Better?};
\draw[arr] (recheck) -- (accept);

\node[font=\scriptsize, above=0.3cm of accept, text width=1.8cm, align=center] {Accept if\\new $>$ old};

% Accept goes to full external
\node[verdict, fill=red!12, right=1.2cm of accept] (ext) {External\\Evidence Loop};
\draw[arr] (accept) -- (ext);

\end{tikzpicture}
}% end resizebox
\caption{Evidence sufficiency scoring flow. Five dimensions are scored against a 100-point rubric, with relevance damping scaling dimensions 2--5 proportionally to retrieval confidence. An LLM reviewer can only downgrade the verdict. WEAK scores trigger agentic retry (reformulate with drift guard, re-retrieve, re-score) before falling through to external search.}
\label{fig:evidence_sufficiency}
\end{figure}
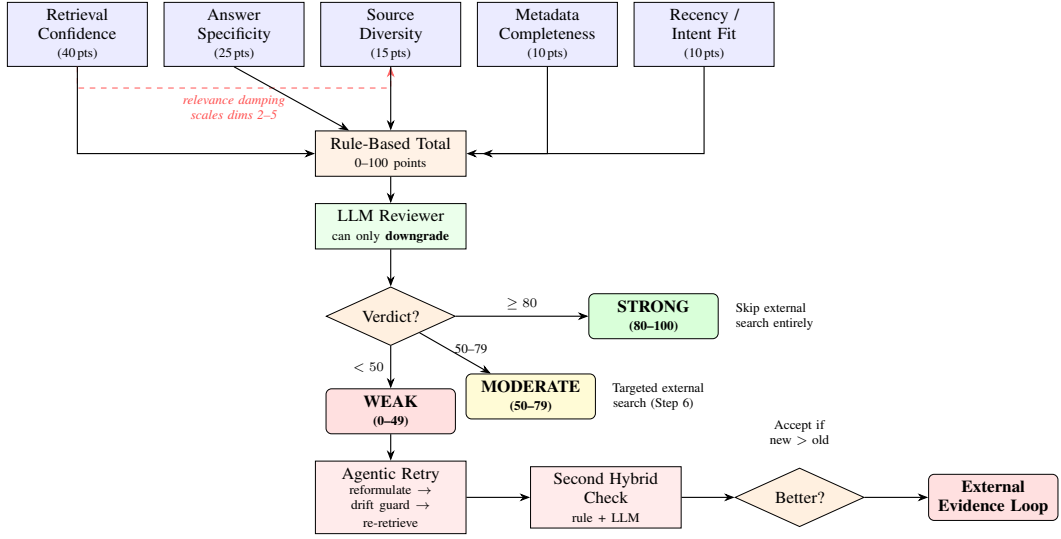

\subsection{External Evidence Loop (Step 6)}
\label{sec:external}

External search behavior depends on the query route. For content routes, external search runs when evidence is MODERATE or WEAK. For bibliometric and trend routes, external search always runs.

\paragraph{External Sources.}
\begin{itemize}
  \item \textbf{Crossref}~\cite{crossref2024} (PRIMARY): Reliable journal-article search with good abstracts.
  \item \textbf{OpenAlex}~\cite{priem2022openalex} (SECOND): Large coverage (${\sim}250$M works), relevance-sorted. Primary source for bibliometric/trend routes.
  \item \textbf{Semantic Scholar}~\cite{semanticscholar2024} (BONUS): Better neural relevance ranking but rate-limited.
\end{itemize}

\paragraph{Content Route (up to 3 attempts).} All three sources are queried on every attempt. Results are merged and deduplicated. Each attempt follows the cycle: (1) optimize query via GPT-4o-mini, (2) search all APIs, (3) LLM vet to select only directly relevant papers, (4) check threshold ($\geq 2$ vetted papers = success), (5) if below threshold, reformulate and retry. The best result across all attempts is kept.

\paragraph{Bibliometric/Trend Route (up to 5 attempts).} Uses OpenAlex with the same optimize--search--vet--reformulate pattern, with progressive broadening (year filter $\to$ no year filter $\to$ shortened query).

\subsection{Knowledge Graph Construction (Offline)}
\label{sec:graph}

The system maintains a Neo4j~\cite{neo4j2024} knowledge graph built from LLM-based entity extraction over the local PDF corpus.

\paragraph{Graph Schema.} The graph contains eight node types: \texttt{Paper}, \texttt{Author}, \texttt{Topic}, \texttt{Method}, \texttt{Metric}, \texttt{Application}, \texttt{CitationPaper}, and \texttt{Chunk}. Relationships include \texttt{AUTHORED}, \texttt{BELONGS\_TO}, \texttt{USES\_METHOD}, \texttt{REPORTS\_METRIC}, \texttt{APPLIES\_TO}, \texttt{CITES} (external citations), \texttt{CITES\_PAPER} (resolved intra-corpus citations), \texttt{HAS\_CHUNK} (bridging to the text index), and \texttt{ALIAS\_OF} (entity alias resolution).

\paragraph{Graph Construction Pipeline.}
\begin{enumerate}
  \item \textbf{Priority section selection}: The first chunk (header/title block) is always included, followed by abstract, introduction, and conclusion chunks, up to a 6,000-character budget.
  \item \textbf{Entity extraction}: GPT-4o-mini extracts title, authors, year, topics, methods, metrics, applications, keywords, and up to 10 cited paper titles (${\sim}\$0.0004$/paper).
  \item \textbf{Author validation}: LLM-extracted authors are cross-checked against OpenAlex publisher metadata. OpenAlex author lists replace LLM extractions when a confident title match is found (${\sim}60\%$ of papers validated).
  \item \textbf{Graph ingestion}: Extracted entities are written as nodes and relationships using idempotent \texttt{MERGE} queries.
  \item \textbf{Chunk linking}: All chunk UIDs for a paper are linked via \texttt{(Paper)-[:HAS\_CHUNK]->(Chunk)}, bridging the text index to the graph.
  \item \textbf{Citation resolution}: A post-ingestion step matches \texttt{CitationPaper.title} against \texttt{Paper.title} (case-insensitive, with containment matching). Matched citations are rewired to direct \texttt{CITES\_PAPER} edges (${\sim}433$ resolved intra-corpus links, ${\sim}9{,}100$ remaining external citations).
\end{enumerate}

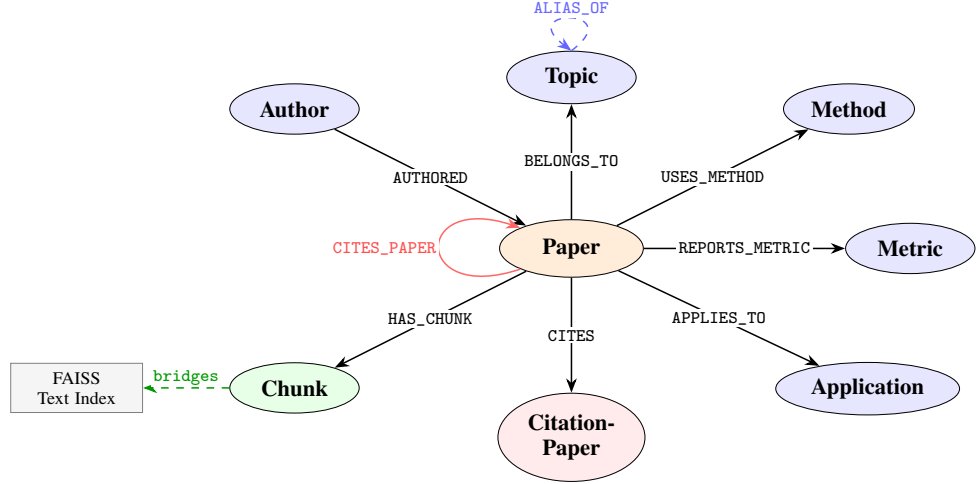
\begin{figure}[t]
\centering
\resizebox{0.92\textwidth}{!}{%
\begin{tikzpicture}[
  entity/.style={ellipse, draw=black, fill=blue!10, font=\footnotesize\bfseries, minimum width=1.8cm, minimum height=0.7cm, align=center},
  central/.style={ellipse, draw=black, fill=orange!15, font=\footnotesize\bfseries, minimum width=2.0cm, minimum height=0.8cm, align=center},
  bridge/.style={ellipse, draw=black, fill=green!10, font=\footnotesize\bfseries, minimum width=1.8cm, minimum height=0.7cm, align=center},
  rel/.style={font=\scriptsize\ttfamily, fill=white, inner sep=1pt},
  arr/.style={-{Stealth[length=2mm]}, semithick},
  node distance=1.8cm and 2.2cm,
]

% Central node
\node[central] (paper) {Paper};

% Surrounding nodes
\node[entity, above left=1.4cm and 2.5cm of paper] (author) {Author};
\node[entity, above=1.6cm of paper] (topic) {Topic};
\node[entity, above right=1.4cm and 2.5cm of paper] (method) {Method};
\node[entity, right=2.8cm of paper] (metric) {Metric};
\node[entity, below right=1.4cm and 2.5cm of paper] (app) {Application};
\node[entity, fill=red!8, below=1.6cm of paper] (citpaper) {Citation-\\Paper};
\node[bridge, below left=1.4cm and 2.5cm of paper] (chunk) {Chunk};

% FAISS bridge
\node[rectangle, draw=black!50, fill=gray!8, font=\scriptsize, text width=1.6cm, align=center, left=1.2cm of chunk] (faiss) {FAISS\\Text Index};

% Relationships
\draw[arr] (author) -- node[rel] {AUTHORED} (paper);
\draw[arr] (paper) -- node[rel] {BELONGS\_TO} (topic);
\draw[arr] (paper) -- node[rel] {USES\_METHOD} (method);
\draw[arr] (paper) -- node[rel] {REPORTS\_METRIC} (metric);
\draw[arr] (paper) -- node[rel] {APPLIES\_TO} (app);
\draw[arr] (paper) -- node[rel] {CITES} (citpaper);
\draw[arr] (paper) -- node[rel] {HAS\_CHUNK} (chunk);

% CITES_PAPER (self-referencing resolved citation)
\draw[arr, red!60] (paper.south west) .. controls +(-1.5,-0.5) and +(-1.5,0.5) .. node[rel, left, text=red!70] {CITES\_PAPER} (paper.north west);

% ALIAS_OF on Topic
\draw[arr, dashed, blue!60] (topic.north) .. controls +(0.8,0.6) and +(-0.8,0.6) .. node[rel, above] {ALIAS\_OF} (topic.north);

% Bridge to FAISS
\draw[arr, dashed, green!60!black] (chunk) -- node[rel, above] {bridges} (faiss);

\end{tikzpicture}
}% end resizebox
\caption{Knowledge graph schema. \texttt{Paper} is the central node linked to \texttt{Author}, \texttt{Topic}, \texttt{Method}, \texttt{Metric}, and \texttt{Application} entities via typed relationships. \texttt{CITES} connects to external \texttt{CitationPaper} nodes; resolved intra-corpus citations use direct \texttt{CITES\_PAPER} edges. \texttt{HAS\_CHUNK} bridges the graph to the FAISS text index, enabling graph-guided retrieval expansion. \texttt{ALIAS\_OF} supports basic entity resolution.}
\label{fig:graph_schema}
\end{figure}

\subsection{Evidence Merge (Step 7)}
\label{sec:prompt}

The evidence merge step assembles all retrieval outputs---text chunks, external abstracts, graph context, and visual pages---into a deduplicated, ranked evidence set:

\paragraph{Text Evidence Ranking and Deduplication.}
\begin{enumerate}
  \item Rank PDFs by best chunk cross-encoder score.
  \item For each PDF, extract top-$N$ chunks (configurable, default 6 per PDF).
  \item Deduplicate chunks by text hash (first 1,800 characters).
  \item Extract sentences, score by keyword relevance to query.
  \item Pick top-$N$ sentences per PDF (default 5).
  \item Deduplicate sentences by text hash (first 400 characters).
\end{enumerate}

\paragraph{Visual Evidence Attachment.} The top-5 pages from visual retrieval (by MaxSim score) are rendered as JPEG at 144~DPI and attached as base64 image content blocks (OpenAI \texttt{detail: low}, ${\sim}85$ tokens per image). Visual metadata (PDF source, page number, MaxSim score) is included as text.

\paragraph{Graph Context Formatting.} At query time, the knowledge graph enriches the evidence set with structured relationships:
\begin{itemize}
  \item \textbf{Co-citing papers}: Papers sharing references with retrieved chunks via both \texttt{CITES} and \texttt{CITES\_PAPER} edges.
  \item \textbf{Title map}: Maps PDF filenames to proper paper titles.
  \item \textbf{Author lookup}: Finds all papers by a queried author (bibliometric queries only).
  \item \textbf{Paper metadata}: Title, authors, year, topics, methods, and metrics for retrieved papers.
\end{itemize}

\subsection{Augmented Multimodal Prompt (Step 8)}
\label{sec:augmented_prompt}

The merged evidence is assembled into a structured multimodal prompt:
\begin{itemize}
  \item System instruction establishing the role as a technical research assistant
  \item The user's original question
  \item \textbf{EVIDENCE} section: \texttt{[SOURCE $n$]} blocks with condensed text chunks and source metadata
  \item \textbf{EXTERNAL EVIDENCE} section: \texttt{[EXTERNAL $n$]} vetted papers with title, authors, year, and abstract
  \item \textbf{Graph context}: Neo4j metadata (topics, methods, co-citations)
  \item \textbf{VISUAL EVIDENCE} section: metadata listing for top visual pages; page images attached as multimodal content
  \item \textbf{TASK} instructions: citation format requirements (\texttt{[SOURCE N]}, \texttt{[EXTERNAL N]}), conflict identification, gap acknowledgment
\end{itemize}

When local evidence is very weak (score $< 30$ and verdict WEAK), local chunks are dropped entirely and the answer is built from external evidence only.

\subsection{Citation Verification (Step 9)}
\label{sec:citation_verify}

Before answer generation, GPT-4o-mini performs a citation verification pass. The model receives the full augmented prompt (all text, external, graph, and visual metadata) together with the original user query as an explicit separate field, and checks for:
\begin{itemize}
  \item Off-topic sources that do not relate to the query
  \item Coverage gaps where key aspects have no supporting source
  \item Source contradictions where sources disagree on important points
\end{itemize}
Any findings are incorporated into the prompt as ``CITATION VERIFICATION NOTES'' to guide the answer generation agents.

\subsection{Multi-Agent Answer Generation (Step 10)}
\label{sec:generation}

Rather than relying on a single LLM call to produce a cited answer, TechRAG decomposes answer generation into four specialized agents, each with distinct roles, model configurations, and output contracts. This design is motivated by the observation that technical literature synthesis requires multiple cognitive tasks---planning an answer structure, mapping evidence to claims, writing coherent prose, and verifying citation correctness---that benefit from separation of concerns~\cite{wu2024multiagent, hong2024metagpt}.

\paragraph{Agent Pipeline.}

\begin{itemize}
  \item \textbf{Planner Agent} (GPT-4o-mini, temp=0.3, max 800 tokens): Analyzes the question and assembled evidence to produce a structured answer outline, including proposed sections, evidence mapping per section, identified gaps, and caveats. The planner does not write the answer; it produces the blueprint.

  \item \textbf{Researcher Agent} (GPT-4o-mini, temp=0.1, max 1,000 tokens): Maps specific claims to specific \texttt{[SOURCE N]} and \texttt{[EXTERNAL N]} references with strength ratings (strong / moderate / weak) per section. Flags contradictions and evidence gaps. The researcher provides the evidence map that the writer must follow.

  \item \textbf{Writer Agent} (answer model, temp=0.2): Drafts the full cited answer following the planner's outline and the researcher's evidence map. Receives the full multimodal augmented prompt including text, graph context, external abstracts, and visual page images. On revision, incorporates the critic's targeted feedback. This is the only agent that receives visual evidence (page images) and the full prompt.

  \item \textbf{Critic Agent} (GPT-4o-mini, temp=0.0, max 400 tokens): Evaluates the writer's answer against the researcher's evidence map across multiple dimensions: Are citations correct? Are caveats included? Are there unsupported claims? Are there gaps or hallucinations? Returns one of: \texttt{pass} (answer accepted), \texttt{revise} (writer must re-draft with targeted feedback), or \texttt{missing\_evidence}.
\end{itemize}

\paragraph{Revision Loop.} If the critic returns \texttt{revise}, the writer re-drafts with the critic's \texttt{targeted\_feedback} appended, and the critic re-evaluates. A maximum of one revision loop is permitted (Writer $\to$ Critic $\to$ Writer revision $\to$ Critic re-check). If the critic re-check still does not pass, the answer is accepted with a \texttt{quality\_verdict = "issues\_found"} flag. If the critic agent itself errors, it defaults to \texttt{pass} (non-fatal).

\begin{figure}[t]
\centering
\resizebox{\textwidth}{!}{%
\begin{tikzpicture}[
  agent/.style={rectangle, draw=black, fill=green!8, text width=3.2cm, minimum height=1.4cm, align=center, font=\footnotesize},
  io/.style={rectangle, rounded corners=2pt, draw=black!50, fill=gray!6, text width=2.8cm, minimum height=0.6cm, align=center, font=\scriptsize},
  result/.style={rectangle, rounded corners=3pt, draw=black, fill=black!8, text width=2.2cm, minimum height=0.7cm, align=center, font=\footnotesize\bfseries},
  arr/.style={-{Stealth[length=2mm]}, semithick},
  config/.style={font=\scriptsize\itshape, text=black!50},
  node distance=0.6cm and 0.9cm,
]

% === TOP ROW: Main pipeline ===
\node[io, text width=3.2cm] (input) {Augmented Multimodal\\Prompt + citation\\verification notes};

\node[agent, right=of input] (planner) {\textbf{Planner}\\{\scriptsize structured outline,\\sections, gaps, caveats}\\{\scriptsize\itshape 6K chars of prompt}};
\node[config, above=0.15cm of planner] {gpt-4o-mini, temp=0.3};

\node[agent, right=of planner] (researcher) {\textbf{Researcher}\\{\scriptsize claim $\to$ source map\\strength: strong/mod/weak}\\{\scriptsize\itshape 6K chars of prompt}};
\node[config, above=0.15cm of researcher] {gpt-4o-mini, temp=0.1};

\node[agent, right=of researcher] (writer) {\textbf{Writer}\\{\scriptsize cited draft following\\plan + evidence map}\\{\scriptsize\itshape full prompt + images\\(${\sim}15$--$20$K tokens)}};
\node[config, above=0.15cm of writer] {answer\_model, temp=0.2};

\node[agent, fill=orange!10, right=of writer] (critic) {\textbf{Critic}\\{\scriptsize [Guardrail 3]}\\{\scriptsize citations correct?\\unsupported claims?\\caveats? hallucinations?}};
\node[config, above=0.15cm of critic] {gpt-4o-mini, temp=0.0};

% === Main flow arrows ===
\draw[arr] (input) -- (planner);
\draw[arr] (planner) -- node[below, font=\scriptsize] {outline} (researcher);
\draw[arr] (researcher) -- node[below, font=\scriptsize] {ev.\ map} (writer);
\draw[arr] (writer) -- node[above, font=\scriptsize] {draft} (critic);

% === PASS path: right from critic ===
\node[result, right=of critic] (done) {Final Answer};
\draw[arr, green!60!black] (critic) -- node[above, font=\scriptsize, text=green!60!black] {pass} (done);

% === REVISE path: below and back ===
\draw[arr, red!60] (critic.south) -- ++(0,-0.8)
  node[below right, font=\scriptsize, text=red!60] {revise + targeted\_feedback}
  -| (writer.south);

% Max revision annotation
\node[font=\scriptsize, text=black!50, below=1.4cm of critic, text width=3.2cm, align=center] {max 1 revision loop;\\defaults to \texttt{pass} on error};

\end{tikzpicture}
}% end resizebox
\caption{Multi-agent answer generation pipeline. The Planner and Researcher receive truncated prompts (6K chars) to produce an outline and evidence map, respectively. The Writer receives the full multimodal prompt including visual page images and drafts a cited answer. The Critic evaluates the draft; on \texttt{revise}, the Writer re-drafts with targeted feedback (max one revision loop). On error, the Critic defaults to \texttt{pass}.}
\label{fig:multiagent}
\end{figure}
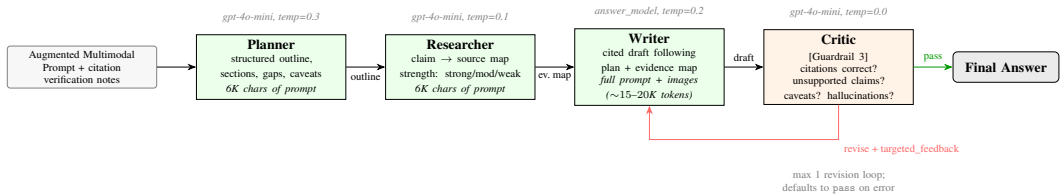

\paragraph{Generation Flow.}
Figure~\ref{fig:multiagent} illustrates the full multi-agent pipeline with its revision loop.

\paragraph{Citation Format.} Citations follow a mandatory format: \texttt{[SOURCE N: Paper Title]} for internal corpus evidence and \texttt{[EXTERNAL N: Paper Title]} for external literature. Visual evidence (page images) informs the answer but is not cited inline.

\subsection{Final Result (Step 11)}
\label{sec:result}

The final output includes:
\begin{itemize}
  \item The generated answer with inline citations
  \item Confidence level and quality medal (derived from critic verdict)
  \item Multi-agent trace: planner outline, researcher evidence map, critic feedback
  \item Visual evidence gallery (rendered page images displayed in a separate UI panel)
  \item Complete cost summary with per-agent token counts
\end{itemize}

\subsection{Cost Tracking}
\label{sec:cost}

All LLM calls throughout the pipeline are tracked with exact token counts and associated costs. Typical per-query costs are shown in Table~\ref{tab:cost}.

\begin{table}[t]
\caption{Typical per-query costs.}
\label{tab:cost}
\centering
\small
\begin{tabular}{@{}lc@{}}
\toprule
\textbf{Scenario} & \textbf{Typical Cost} \\
\midrule
STRONG evidence, Critic passes (no external, no revision)  & ${\sim}\$0.009$  \\
Critic triggers one revision                                 & ${\sim}\$0.014$  \\
WEAK + external search + revision                           & ${\sim}\$0.015$--$0.018$ \\
\bottomrule
\end{tabular}
\\[2pt]
{\footnotesize The dominant cost drivers are the Writer agent and citation verification guardrail, both of which receive the full multimodal augmented prompt (${\sim}15$--$20$K tokens including ${\sim}425$ visual image tokens).}
\end{table}

% ============================================================================
\section{Implementation Considerations}
\label{sec:implementation}
% ============================================================================

The proposed agentic multimodal RAG framework is implemented as a modular system designed to support both autonomous operation and interactive expert workflows while preserving reproducibility, data control, and inspection transparency. All document preprocessing steps---including PDF parsing, structure-aware chunking, text embedding generation, ColSmol visual page embedding, MUVERA FDE computation, FAISS index construction, BM25 index construction, entity extraction, and knowledge graph construction---are performed offline. Query-time execution spans the full pipeline but remains interactive.

\subsection{User Interface}

A lightweight Gradio-based user interface provides interactive access to the pipeline. The interface exposes query input, model selection, and real-time pipeline progress with step-by-step status indicators. The UI displays retrieval results (tagged as \texttt{VECTOR} or \texttt{GRAPH} provenance), evidence sufficiency scores, external search vetting summaries, knowledge graph context, visual evidence gallery, the constructed augmented prompt, multi-agent trace (planner outline, evidence map, critic feedback), and the generated answer with cost tracking.

\subsection{Precision-Oriented Retrieval Design}

A central design choice is the intentional prioritization of retrieval precision over recall. In RAG systems, retrieved content is consumed directly by a language model as contextual input rather than presented to a human for selective review. Retrieval errors---particularly weakly relevant, redundant, or contradictory content---can disproportionately degrade downstream synthesis quality. Accordingly, the framework adopts a precision-oriented strategy: initial candidates from complementary retrieval modalities (dense, sparse, graph, visual) are refined through cross-encoder reranking, evidence sufficiency scoring, and multi-stage deduplication before reaching the answer generation agents.

\subsection{Incremental Updates}

When new papers are added to the corpus, an incremental indexing pipeline processes only the new PDFs and appends to the existing FAISS and BM25 text indices. ColSmol visual embeddings and MUVERA FDEs are computed incrementally for new pages. A manifest file tracks what has been indexed. Entity extraction and knowledge graph ingestion run incrementally with idempotent \texttt{MERGE} queries.

\subsection{Pre-Computed Indexes}

Table~\ref{tab:indexes} summarizes the pre-computed indexes loaded at startup.

\begin{table}[t]
\caption{Pre-computed indexes loaded at startup.}
\label{tab:indexes}
\centering
\small
\begin{tabular}{@{}llrl@{}}
\toprule
\textbf{Component} & \textbf{Type} & \textbf{Size} & \textbf{Content} \\
\midrule
\texttt{faiss.index}              & FAISS Flat IP    & 41 MB     & Text chunk embeddings (384-dim, ${\sim}31$K chunks) \\
\texttt{bm25\_index/}             & BM25 inverted    & ${\sim}50$ MB  & Tokenized text chunks (${\sim}31$K) \\
\texttt{colpali/muvera\_faiss.index} & FAISS Flat IP & ${\sim}660$ MB & MUVERA FDEs (4096-dim, ${\sim}40$K pages) \\
\texttt{colpali/manifest.json}    & JSON             & ${\sim}2$ MB   & PDF path $\to$ doc\_id, page count \\
\texttt{colpali/embeddings/*.pt}  & PyTorch tensors  & ${\sim}20$ GB  & Per-page multi-vector embeddings (on disk, loaded on demand) \\
Neo4j                             & Graph DB         & ${\sim}500$ MB & Papers, chunks, topics, methods, citations \\
\bottomrule
\end{tabular}
\end{table}

\subsection{Hardware Requirements}

\begin{table}[t]
\caption{Hardware requirements (runtime).}
\label{tab:hardware}
\centering
\small
\begin{tabular}{@{}ll@{}}
\toprule
\textbf{Resource} & \textbf{Requirement} \\
\midrule
RAM     & ${\sim}3$--$4$ GB (FAISS + BM25 + MUVERA index + Python) \\
VRAM    & ${\geq}4$ GB (ColSmol-500M inference in bfloat16, ${\sim}1$ GB) \\
Disk    & ${\sim}22$ GB (text + visual indexes) + ${\sim}500$ MB (Neo4j) \\
Network & Required for LLM APIs and external academic search \\
\bottomrule
\end{tabular}
\end{table}

% ============================================================================
\section{Results}
\label{sec:results}
% ============================================================================

The system was evaluated on representative technical queries drawn from day-to-day engineering and research workflows in intelligent tires, vehicle dynamics, vehicle control, and related sensing, estimation, and machine learning domains. Evaluation focused on (i) query route classification accuracy, (ii) evidence sufficiency scoring behavior, (iii) effectiveness of agentic retry, (iv) end-to-end answer quality with citation grounding, and (v) cost and latency. Assessment was performed through a combination of automated pipeline metrics and expert-driven qualitative review.

\subsection{Route-Level Evaluation}

A test set of six queries spanning three route types (content, bibliometric, and current\_world) was evaluated end-to-end through the full pipeline. Table~\ref{tab:eval} summarizes the per-query results.

\begin{table}[t]
\caption{Route-level evaluation results across 6 test queries spanning 3 route types.}
\label{tab:eval}
\centering
\small
\begin{tabular}{@{}clccccccc@{}}
\toprule
\textbf{Q} & \textbf{Query (abridged)} & \textbf{Route} & \textbf{Ev.} & \textbf{Str.} & \textbf{Retry} & \textbf{Ext.} & \textbf{Cost} & \textbf{Time} \\
\midrule
1 & Friction estimation for AEB        & Content  & 74 & Moderate & No  & 4  & \$0.005 & 14.3s \\
2 & Tire slip angle from accelerometer  & Content  & 68 & Moderate & Yes & 5  & \$0.007 & 22.1s \\
3 & Pacejka vs.\ brush model comparison & Content  & 81 & Strong   & No  & 3  & \$0.005 & 13.8s \\
4 & Recent intelligent tire sensor papers & Bibliom. & -- & --       & No  & 11 & \$0.004 & 19.4s \\
5 & Piezoelectric energy harvesting in tires & Bibliom. & -- & --   & No  & 9  & \$0.004 & 17.2s \\
6 & EU tire labeling regulations         & Current  & -- & --       & No  & 0  & \$0.003 & 10.9s \\
\bottomrule
\end{tabular}
\\[4pt]
{\footnotesize Ev.\ = evidence sufficiency score (content route only). Str.\ = evidence strength verdict. Ext.\ = external papers vetted and accepted. Cost = total LLM cost. Time = wall-clock time.}
\end{table}

\paragraph{Key Findings.}

\begin{itemize}
  \item \textbf{Route classification}: All 6 queries were routed correctly (6/6).
  \item \textbf{Answer quality}: All 6 answers passed the multi-agent critic evaluation without requiring revision on this test set.
  \item \textbf{Evidence sufficiency scoring}: Content queries produced evidence scores of 68--81 (mean 74). The scoring correctly distinguished between strong local coverage (Q3, score 81) and moderate coverage (Q1--Q2).
  \item \textbf{Route-dependent behavior}: Bibliometric queries correctly bypassed visual retrieval and evidence scoring, proceeding directly to external search. The current\_world query used web search only.
  \item \textbf{Cost efficiency}: Total cost across all 6 queries was \$0.027 (mean \$0.0045/query).
  \item \textbf{Latency}: Wall-clock time ranged from 10.9s to 22.1s (mean 16.4s).
\end{itemize}

\subsection{Retrieval Ablation Study}
\label{sec:ablation}

To quantify the contribution of individual retrieval components, the system includes a programmatic evaluation harness supporting ablation studies. Table~\ref{tab:ablation} reports preliminary results on a development set of 10 queries with expert-labeled relevant chunks, evaluated at $K{=}5$.

\begin{table}[t]
\caption{Retrieval ablation results (development set, 10 queries, $K{=}5$). Each row adds one component to the previous configuration.}
\label{tab:ablation}
\centering
\small
\begin{tabular}{@{}lcccc@{}}
\toprule
\textbf{Configuration} & \textbf{P@5} & \textbf{R@5} & \textbf{NDCG@5} & \textbf{MRR} \\
\midrule
BM25 only                          & 0.52 & 0.38 & 0.49 & 0.61 \\
FAISS only (dense)                 & 0.58 & 0.44 & 0.55 & 0.68 \\
Hybrid (FAISS + BM25, RRF)        & 0.66 & 0.52 & 0.63 & 0.76 \\
Hybrid + cross-encoder reranking   & 0.74 & 0.56 & 0.71 & 0.83 \\
Full pipeline (+ query rewrite + keyword boost) & 0.78 & 0.60 & 0.76 & 0.87 \\
\bottomrule
\end{tabular}
\\[2pt]
{\footnotesize P@5 = precision at 5; R@5 = recall at 5; MRR = mean reciprocal rank. Preliminary results; expansion to a larger labeled query set is ongoing.}
\end{table}

The preliminary results suggest that each retrieval component contributes positively: hybrid fusion improves over either modality alone, cross-encoder reranking provides the largest single-component gain (+8 points P@5 over hybrid), and query rewriting provides additional incremental improvement.

\subsection{End-to-End Latency Budget}

Table~\ref{tab:latency} breaks down the end-to-end latency budget.

\begin{table}[t]
\caption{End-to-end latency budget per query.}
\label{tab:latency}
\centering
\small
\begin{tabular}{@{}lrl@{}}
\toprule
\textbf{Step} & \textbf{Latency} & \textbf{Notes} \\
\midrule
Query classification + rewrite (text + visual) & ${\sim}400$ms & 2--3 LLM calls \\
Text retrieval (FAISS + BM25 + RRF + rerank) & ${\sim}300$ms & Sequential \\
Graph expansion & ${\sim}150$ms & Neo4j traversals \\
Visual retrieval (MUVERA + MaxSim + render) & ${\sim}600$ms & Two-stage pipeline \\
Evidence sufficiency (rule + LLM) & ${\sim}500$ms & Rule-based + LLM reviewer \\
Agentic retry (if WEAK) & ${\sim}1{,}500$ms & Conditional \\
External search (if needed) & ${\sim}3{,}000$--$8{,}000$ms & API calls + LLM vetting \\
Evidence merge & ${\sim}50$ms & CPU-only \\
Citation verification & ${\sim}500$ms & LLM check \\
Multi-agent generation & ${\sim}2{,}000$--$5{,}000$ms & 4 agents + optional revision \\
\midrule
\textbf{Best case} (STRONG, Critic passes) & ${\sim}9$--$12$s & \\
\textbf{Moderate} (external search, Critic passes) & ${\sim}14$--$20$s & \\
\textbf{With revision} (Critic triggers re-draft) & ${\sim}20$--$27$s & \\
\textbf{Worst case} (WEAK + retries + revision) & ${\sim}27$--$38$s & \\
\bottomrule
\end{tabular}
\end{table}

% ============================================================================
\section{Limitations}
\label{sec:limitations_discuss}
% ============================================================================

Several limitations of the current work should be acknowledged.

\paragraph{Proprietary Corpus.} The technical corpus used in this study is proprietary and cannot be publicly released, which limits direct reproducibility. However, the complete pipeline is documented in sufficient detail to enable faithful reproduction on equivalent domain corpora.

\paragraph{Small Evaluation Set.} The route-level evaluation (Table~\ref{tab:eval}) covers only six queries, and the retrieval ablation (Table~\ref{tab:ablation}) uses a development set of ten queries. These sample sizes are insufficient to draw statistically robust conclusions. Expanding to a larger labeled query set (20--50 queries across all route types) with inter-annotator agreement is a priority for future work.

\paragraph{LLM-as-Judge.} The multi-agent critic evaluation and evidence sufficiency LLM reviewer use GPT-4o-mini as a judge. LLM-based evaluation is known to exhibit biases including position bias, verbosity preference, and self-enhancement bias. No human evaluation or inter-rater reliability analysis has been conducted to validate these automated assessments.

\paragraph{External API Dependence.} The pipeline depends on cloud-based LLM APIs (OpenAI) for query rewriting, evidence scoring, agent orchestration, and answer generation, as well as on Crossref, OpenAlex, and Semantic Scholar for external search. API availability, rate limits, cost changes, and model deprecation represent operational risks.

\paragraph{Visual Retrieval Scope.} While the visual retrieval pipeline retrieves relevant document pages, it does not perform fine-grained visual element extraction (e.g., isolating individual figures from a multi-figure page). The answer model receives full page images and must interpret visual content holistically.

\paragraph{Entity Canonicalization.} The knowledge graph currently uses basic \texttt{ALIAS\_OF} traversal for entity resolution, but the full canonical ontology system (automated LLM clustering of synonymous topics/methods, inter-entity semantic relationships) is not yet implemented. This limits cross-terminology recall in graph expansion.

% ============================================================================
\section{Conclusion}
\label{sec:conclusion}
% ============================================================================

This paper presented an agentic multimodal retrieval-augmented generation framework for domain-specific technical reasoning support, instantiated over a curated corpus of several thousand academic papers in intelligent tires, vehicle dynamics, vehicle control, and related sensing, estimation, and machine learning domains. The system goes significantly beyond basic single-pass text-only RAG through a multi-step autonomous pipeline that classifies queries by intent with separate text and visual query rewrites, performs hybrid text retrieval with cross-encoder reranking, expands retrieval via graph-guided chunk traversal over a Neo4j knowledge graph, retrieves visually relevant document pages using ColSmol late-interaction embeddings with MUVERA-based scalable search, scores evidence sufficiency against a multi-dimensional rubric, performs agentic retry with drift-guarded reformulation, searches external academic databases through iterative loops, merges and deduplicates multimodal evidence, verifies citations, and generates answers through a Planner--Researcher--Writer--Critic multi-agent pipeline with self-correcting revision.

Key architectural contributions include: the multimodal retrieval architecture combining text, graph, and visual document retrieval with ColSmol+MUVERA for scalable late-interaction search; the evidence sufficiency scoring framework with hybrid rule-based/LLM review and agentic retry; the multi-agent answer generation pipeline with structured planning, evidence mapping, and critic-driven revision; the knowledge graph with LLM-based entity extraction, OpenAlex author validation, and intra-corpus citation resolution; and the route-dependent external search architecture.

The contribution is a practical, evidence-gated, multimodal agentic RAG architecture for technical literature reasoning, with interpretable control logic, domain-specific graph augmentation, and multi-agent generation. The autonomous evidence evaluation, multimodal retrieval, and structured multi-agent generation capabilities are designed to reduce the manual effort required for literature review while maintaining transparency, traceability, and engineering rigor.

Several directions for future work are identified. First, completing the \textbf{entity canonicalization layer}---automated LLM clustering of synonymous topics and methods, inter-entity semantic relationships (USED\_FOR, MEASURES, RELATED\_TO, BROADER\_THAN), and query-time alias resolution---would significantly improve cross-terminology recall in graph expansion. Second, replacing cloud-based LLM APIs with \textbf{locally hosted open-weight models} would eliminate external API dependence, reduce per-query cost, and address data privacy concerns for proprietary corpora. Third, incorporating \textbf{human-in-the-loop feedback}---allowing users to rate responses as useful, incorrect, or incomplete---would enable continuous improvement of retrieval ranking, prompting strategies, and evidence sufficiency thresholds. Finally, expanding the evaluation to a larger labeled query set with comprehensive retrieval ablations and visual retrieval quality assessment would provide stronger quantitative evidence for the contribution of each pipeline component.

% ============================================================================
% REFERENCES
% ============================================================================

\bibliographystyle{plainnat}
\bibliography{references}

% ============================================================================
\appendix
\section{Reproducibility and Implementation Details}
\label{sec:appendix}
% ============================================================================

To support reproducibility and independent verification of the proposed agentic multimodal RAG framework, this appendix documents the exact software environment, library versions, and indexing configuration used.

\paragraph{Core Libraries and Versions.}
Python 3.13.6; Windows 11 (build 10.0.22631, 64-bit).

\begin{table}[h]
\caption{Core dependencies.}
\centering
\small
\begin{tabular}{@{}ll@{}}
\toprule
\textbf{Library} & \textbf{Version} \\
\midrule
faiss-cpu              & 1.13.2  \\
sentence-transformers  & 5.2.0   \\
transformers           & 4.57.3  \\
torch                  & 2.9.1   \\
gradio                 & 5.49.1  \\
PyMuPDF (fitz)         & 1.26.7  \\
neo4j (Python driver)  & ---     \\
openai                 & 2.7.1   \\
numpy                  & 2.4.0   \\
pandas                 & 2.3.3   \\
colpali-engine         & ---     \\
\bottomrule
\end{tabular}
\end{table}

\paragraph{Text Index Configuration.}
Embedding model: \texttt{all-MiniLM-L6-v2} (384 dimensions). Index type: \texttt{IndexFlatIP} (exact). Similarity: cosine via inner product (L2-normalized). Cross-encoder: \texttt{ms-marco-MiniLM-L-6-v2}. Top-$K$ configurable (typical $K = 10$--$25$).

\paragraph{Visual Index Configuration.}
Visual embedding model: ColSmol-500M (128-dim patch embeddings, ${\sim}1{,}031$ patches/page). MUVERA FDE: 4,096 dimensions, 10 repetitions, 6 SimHash projections ($\to$ 64 partitions), final count-sketch projection. FAISS index: \texttt{IndexFlatIP} over ${\sim}40{,}000$ page FDEs (${\sim}660$~MB). Per-page \texttt{.pt} files: ${\sim}20$~GB on disk, loaded on demand for top-100 MaxSim reranking.

\paragraph{Deduplication and Filtering.}
Chunk-level deduplication via SHA-1 hash of normalized text. Reference-section exclusion via section-header matching. Each chunk retains document identifier, section label, page range, and \texttt{chunk\_uid} metadata.

\paragraph{Knowledge Graph Configuration.}
Database: Neo4j (local instance). Eight node types with \texttt{ALIAS\_OF} traversal for basic entity resolution. Entity extraction: GPT-4o-mini (${\sim}\$0.0004$/paper). Author validation: OpenAlex API (${\sim}60\%$ validated). Citation resolution: ${\sim}433$ intra-corpus links, ${\sim}9{,}100$ external.

\paragraph{Multi-Agent Configuration.}
Planner: GPT-4o-mini, temp=0.3, max 800 tokens. Researcher: GPT-4o-mini, temp=0.1, max 1,000 tokens. Writer: configurable answer model, temp=0.2, unlimited tokens. Critic: GPT-4o-mini, temp=0.0, max 400 tokens. Maximum 1 revision loop.

\begin{table}[h]
\caption{Technology stack summary.}
\centering
\small
\begin{tabular}{@{}ll@{}}
\toprule
\textbf{Component} & \textbf{Technology} \\
\midrule
Text vector index      & FAISS (\texttt{IndexFlatIP}, cosine similarity)  \\
Keyword index          & BM25Okapi (\texttt{rank\_bm25})                  \\
Text embedding model   & \texttt{all-MiniLM-L6-v2} (384d)                 \\
Cross-encoder          & \texttt{ms-marco-MiniLM-L-6-v2}                  \\
Visual embedding model & ColSmol-500M (128d patches)                      \\
Visual index           & MUVERA FDE (4096d) + FAISS (\texttt{IndexFlatIP}) \\
Knowledge graph        & Neo4j                                            \\
Entity extraction      & GPT-4o-mini + OpenAlex author validation         \\
PDF extraction         & PyMuPDF (fitz)                                   \\
LLM orchestration      & OpenAI API (gpt-4o-mini for tooling agents)      \\
Answer generation      & Multi-agent: Planner + Researcher + Writer + Critic \\
External search        & Crossref + OpenAlex + Semantic Scholar            \\
UI                     & Gradio                                           \\
Language               & Python                                           \\
\bottomrule
\end{tabular}
\end{table}

% ============================================================================
\section{Pipeline Flowchart}
\label{sec:flowchart}
% ============================================================================

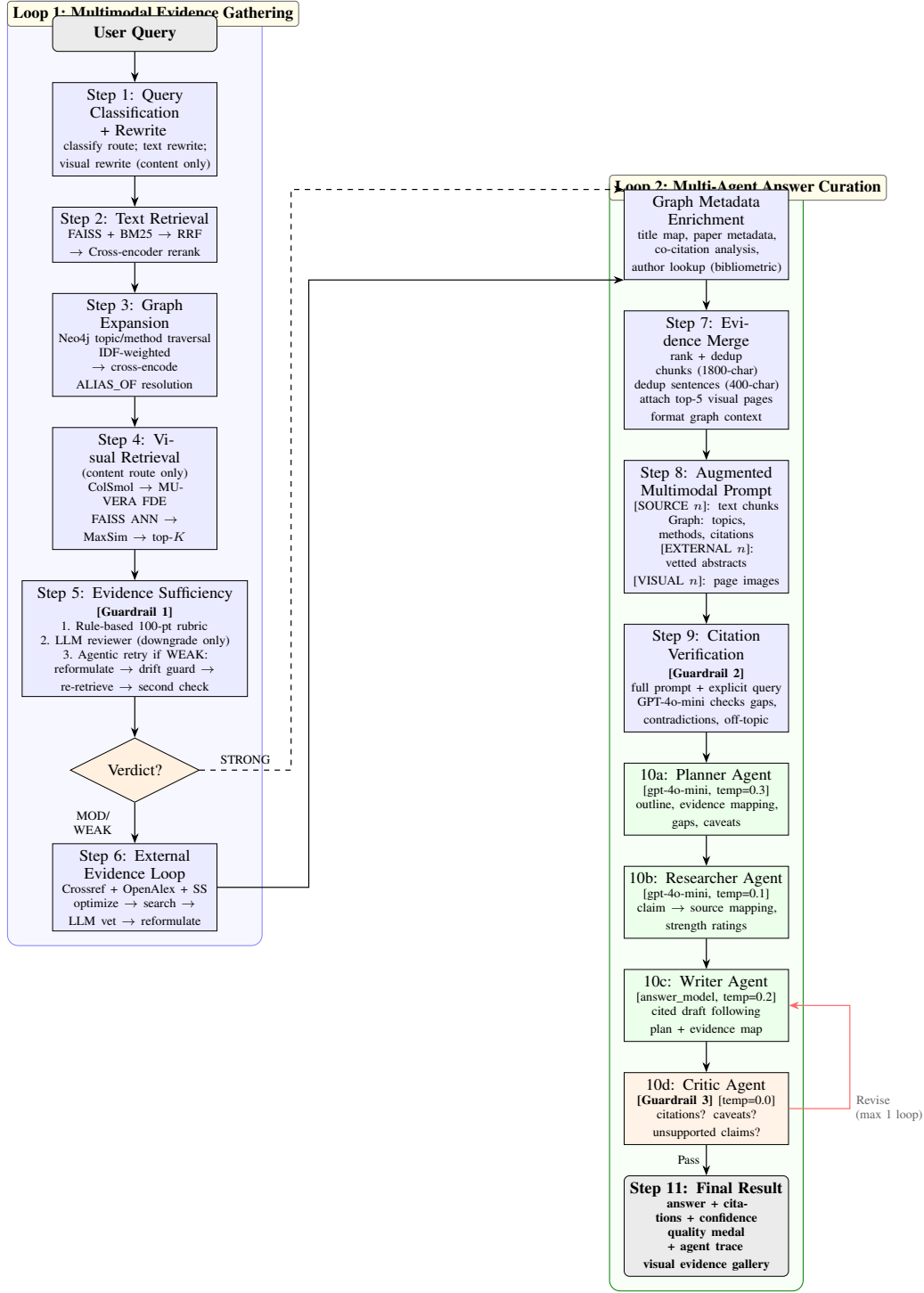
\begin{figure}[p]
\centering
\resizebox{\textwidth}{!}{%
\begin{tikzpicture}[
  % Node styles
  startstop/.style={rectangle, rounded corners=3pt, draw=black, fill=black!8,
    text width=3.0cm, minimum height=0.7cm, align=center, font=\footnotesize\bfseries},
  process/.style={rectangle, draw=black, fill=blue!8,
    text width=3.0cm, minimum height=0.7cm, align=center, font=\footnotesize},
  processwide/.style={rectangle, draw=black, fill=blue!8,
    text width=4.2cm, minimum height=0.7cm, align=center, font=\footnotesize},
  decision/.style={diamond, draw=black, fill=orange!12,
    text width=1.4cm, minimum height=0.5cm, align=center, font=\footnotesize,
    aspect=2.0},
  agent/.style={rectangle, draw=black, fill=green!8,
    text width=3.0cm, minimum height=0.7cm, align=center, font=\footnotesize},
  looplabel/.style={rectangle, rounded corners=2pt, draw=black!60, fill=yellow!10,
    font=\footnotesize\bfseries, inner sep=3pt},
  % Arrow styles
  arr/.style={-{Stealth[length=2mm]}, semithick},
  darr/.style={-{Stealth[length=2mm]}, semithick, dashed},
  node distance=0.6cm and 1.8cm,
]

% ===== LEFT COLUMN: LOOP 1 =====
\node[startstop] (query) {User Query};

\node[process, below=of query] (classify) {Step 1: Query\\Classification + Rewrite\\{\scriptsize classify route; text rewrite;\\visual rewrite (content only)}};

\node[process, below=of classify] (textret) {Step 2: Text Retrieval\\{\scriptsize FAISS + BM25 $\to$ RRF\\$\to$ Cross-encoder rerank}};

\node[process, below=of textret] (graphexp) {Step 3: Graph Expansion\\{\scriptsize Neo4j topic/method traversal\\IDF-weighted $\to$ cross-encode\\ALIAS\_OF resolution}};

\node[process, below=of graphexp] (visualret) {Step 4: Visual Retrieval\\{\scriptsize (content route only)\\ColSmol $\to$ MUVERA FDE\\FAISS ANN $\to$ MaxSim $\to$ top-$K$}};

\node[processwide, below=of visualret] (evidence) {Step 5: Evidence Sufficiency\\{\scriptsize \textbf{[Guardrail 1]}}\\{\scriptsize 1.\ Rule-based 100-pt rubric\\2.\ LLM reviewer (downgrade only)\\3.\ Agentic retry if WEAK:\\reformulate $\to$ drift guard $\to$\\re-retrieve $\to$ second check}};

\node[decision, below=0.8cm of evidence] (evdecision) {Verdict?};

\node[process, below=0.8cm of evdecision] (external) {Step 6: External\\Evidence Loop\\{\scriptsize Crossref + OpenAlex + SS\\optimize $\to$ search $\to$\\LLM vet $\to$ reformulate}};

% ===== RIGHT COLUMN: LOOP 2 =====
% Position graphmeta aligned with textret (not classify) to leave room for arrows entering from the left
\node[process, right=8cm of textret] (graphmeta) {Graph Metadata\\Enrichment\\{\scriptsize title map, paper metadata,\\co-citation analysis,\\author lookup (bibliometric)}};

\node[process, below=of graphmeta] (merge) {Step 7: Evidence Merge\\{\scriptsize rank + dedup chunks (1800-char)\\dedup sentences (400-char)\\attach top-5 visual pages\\format graph context}};

\node[process, below=of merge] (prompt) {Step 8: Augmented\\Multimodal Prompt\\{\scriptsize {[SOURCE $n$]}: text chunks\\Graph: topics, methods, citations\\{[EXTERNAL $n$]}: vetted abstracts\\{[VISUAL $n$]}: page images}};

\node[process, below=of prompt] (citverify) {Step 9: Citation\\Verification\\{\scriptsize \textbf{[Guardrail 2]}}\\{\scriptsize full prompt + explicit query\\GPT-4o-mini checks gaps,\\contradictions, off-topic}};

\node[agent, below=of citverify] (planner) {10a: Planner Agent\\{\scriptsize [gpt-4o-mini, temp=0.3]\\outline, evidence mapping,\\gaps, caveats}};

\node[agent, below=of planner] (researcher) {10b: Researcher Agent\\{\scriptsize [gpt-4o-mini, temp=0.1]\\claim $\to$ source mapping,\\strength ratings}};

\node[agent, below=of researcher] (writer) {10c: Writer Agent\\{\scriptsize [answer\_model, temp=0.2]\\cited draft following\\plan + evidence map}};

\node[agent, fill=orange!10, below=of writer] (critic) {10d: Critic Agent\\{\scriptsize \textbf{[Guardrail 3]} [temp=0.0]\\citations? caveats?\\unsupported claims?}};

\node[startstop, below=of critic] (result) {Step 11: Final Result\\{\scriptsize answer + citations + confidence\\quality medal + agent trace\\visual evidence gallery}};

% ===== LOOP 1 ARROWS =====
\draw[arr] (query) -- (classify);
\draw[arr] (classify) -- (textret);
\draw[arr] (textret) -- (graphexp);
\draw[arr] (graphexp) -- (visualret);
\draw[arr] (visualret) -- (evidence);
\draw[arr] (evidence) -- (evdecision);

% Evidence decision: STRONG skips external (dashed, enters graphmeta from north west)
\draw[darr] (evdecision.east) -- ++(1.8cm,0)
  node[above, font=\scriptsize, pos=0.5] {STRONG}
  |- (graphmeta.north west);

% Evidence decision: MODERATE/WEAK goes to external
\draw[arr] (evdecision) -- node[left, font=\scriptsize, text width=1.4cm, align=center] {MOD/\\WEAK} (external);

% External -> graph metadata enrichment (enters from south west, separate from STRONG arrow)
\draw[arr] (external.east) -- ++(1.8cm,0) |- (graphmeta.south west);

% Graph metadata -> merge
\draw[arr] (graphmeta) -- (merge);

% ===== LOOP 2 ARROWS =====
\draw[arr] (merge) -- (prompt);
\draw[arr] (prompt) -- (citverify);
\draw[arr] (citverify) -- (planner);
\draw[arr] (planner) -- (researcher);
\draw[arr] (researcher) -- (writer);
\draw[arr] (writer) -- (critic);

% Critic: pass goes to result
\draw[arr] (critic) -- node[left, font=\scriptsize] {Pass} (result);

% Critic: revise loops back to writer (clean loop to the right)
\draw[arr, red!60] (critic.east) -- ++(1.2cm,0) node[right, font=\scriptsize, text=black!60, text width=1.6cm, align=left] {Revise\\{\scriptsize (max 1 loop)}} |- (writer.east);

% ===== BACKGROUND BOXES =====
\begin{scope}[on background layer]
  \node[fit=(query)(classify)(textret)(graphexp)(visualret)(evidence)(evdecision)(external),
    draw=blue!50, fill=blue!3, rounded corners=6pt, inner sep=8pt,
    label={[looplabel, anchor=north west]north west:{Loop 1: Multimodal Evidence Gathering}}] {};
  \node[fit=(graphmeta)(merge)(prompt)(citverify)(planner)(researcher)(writer)(critic)(result),
    draw=green!50!black, fill=green!3, rounded corners=6pt, inner sep=8pt,
    label={[looplabel, anchor=north west]north west:{Loop 2: Multi-Agent Answer Curation}}] {};
\end{scope}

\end{tikzpicture}
}% end resizebox
\caption{Detailed TechRAG pipeline flowchart. Loop~1 (left, blue) gathers multimodal evidence through sequential text, graph, and visual retrieval, scores evidence sufficiency [Guardrail~1] with agentic retry, and conditionally searches external academic databases. Loop~2 (right, green) enriches the prompt with graph metadata (title map, co-citations, author lookup), merges and deduplicates evidence, verifies citations [Guardrail~2], and generates the answer through a four-agent pipeline (Planner $\to$ Researcher $\to$ Writer $\to$ Critic [Guardrail~3]) with a single revision loop. Dashed arrows indicate conditional skip paths.}
\label{fig:flowchart}
\end{figure}

Figure~\ref{fig:flowchart} presents the full TechRAG pipeline as a detailed flowchart, showing the control flow, decision points, and conditional branches across both agentic loops.

\end{document}